\documentclass[preprintnumbers,prd,onecolumn,floatfix,superscriptaddress, nofootinbib]{revtex4-2}

\usepackage{setspace}
\usepackage{graphicx}
\usepackage{subfig}
\usepackage{epsfig}
\usepackage{bm}
\usepackage{amssymb}
\usepackage{float}
\usepackage{amsmath}
\usepackage{dcolumn}
\usepackage[colorlinks]{hyperref}
\usepackage[usenames,dvipsnames]{color} \usepackage{enumitem}
\usepackage{orcidlink} 

\usepackage{booktabs}
\usepackage{threeparttable}
\usepackage{multirow}
\usepackage{xcolor}
\definecolor{bestfitblue}{RGB}{0,0,180}

\hypersetup{ breaklinks=true, pdfstartview={FitH}, colorlinks=true, linkcolor=blue, citecolor=red, filecolor=magenta, urlcolor=blue, anchorcolor=green, linktocpage=true }

\def\doi{http://doi.org}

\newcommand{\be}{\begin{equation}}
\newcommand{\ee}{\end{equation}}
\newcommand{\beano}{\begin{Eqarray*}}
\newcommand{\eeano}{\end{Eqarray*}}
\newcommand{\ba}{\begin{Eqarray}}
\newcommand{\ea}{\end{Eqarray}}

\begin{document}

\title{The cosmic consequences and the constraints on HN-gravity}

\author{J. K. Singh\orcidlink{0009-0000-6037-6702}}
\email{jksingh@nsut.ac.in}
\affiliation{Department of Mathematics, Netaji Subhas University of Technology, New Delhi-110078, India}

\author{Sonal Aggarwal\orcidlink{0009-0009-0481-6799}}
\email{sonal.aggarwal@nsut.ac.in}
\affiliation{Department of Mathematics, Netaji Subhas University of Technology, New Delhi-110078, India}

\author{Shaily\orcidlink{0000-0002-9598-5900}}
\email{shailytyagi.iitkgp@gmail.com}
\affiliation{School of Computer Science Engineering and Technology, Bennett University, Greater Noida, India}

\author{Hamid Shabani\orcidlink{0000-0002-2309-3591}}
\email{h.shabani@phys.usb.ac.ir}
\affiliation{Physics Department, Faculty of Sciences, University of Sistan and Baluchestan, Zahedan, Iran}

\author{Joao R. L. Santos\orcidlink{0000-0002-9688-938X}}
\email{joaorafael@df.ufcg.edu.br}
\affiliation{Unidade Acadêmica de F\'{\i}sica, Universidade Federal de Campina Grande,\\ Caixa Postal 10071, 58429-900, Campina Grande, Para\'{\i}ba, Brazil \\
Unidade Acad\^emica de Matem\'atica, Universidade Federal de Campina Grande,\\ 58429-970,  Campina Grande, Para\'{\i}ba, Brazil\\
Institute for Theoretical Physics, Leibniz University Hannover, Appelstraße 2, 30167 Hannover, Germany\\
Institut für Theoretische Physik, Universität Heidelberg, Philosophenweg 16, 69120 Heidelberg, Germany.}

\begin{abstract}

In this paper, we investigate the late-time cosmic acceleration of the Quintessence model within the framework of Hoyle–Narlikar Gravity (HNG), which incorporates a creation field. Using the Hubble tension as a function of the density parameter for matter, the density parameter for radiation, and the density parameter for dark energy in the covariant formulation, we find the gravitational field equations in the spatially flat, homogeneous, and isotropic spacetime to examine the dynamical mechanism that leads to cosmic acceleration in the late-time universe. We analyze the observational constraints on the late-time density parameters using various recent observational datasets, including the Hubble datasets, Pantheon$^{+}$, and joint compilation, Pantheon$^{+}$+BAO. Consequently, it is explicitly demonstrated that late-time cosmic acceleration can be consistent with recent observational data in Hoyle–Narlikar Gravity with non-minimal matter interaction. In contrast with other modified theories of gravity, it is observed that the creation field theory with non-minimal matter interaction renders more compact constraints on the Hubble tension together with density parameters, and extensively explains the accelerating expansion of the universe, which makes it a more plausible option compared to the $ \Lambda $CDM model. Furthermore, the $ w-w' $ phase analysis confirms alternating thawing and freezing behaviour of the model, with all trajectories ultimately converging toward the $ \Lambda $CDM point, thereby confirming the model stability and the observational consistency.

\end{abstract}

\maketitle

\section{Introduction} \label{sec:1}

Einstein's General Relativity in $ 1915 $ \cite{Einstein:1915ca} changed our understanding of gravity and the structure of the Universe forever. This was justified when Edwin Hubble discovered in 1929 that the universe is expanding, providing the first evidence for the Big Bang~\cite{Hubble:1929ig}. The additional confirmation came in $ 1965 $ with the discovery of cosmic microwave background radiation by Arno Penzias and Robert Wilson~\cite{Penzias:1965wn}, which confirmed that model. Initially, it was thought that the expansion of the Universe would be slowly depressed by gravity. However, in the late $ 1990s $, teams led by Saul Perlmutter, Brian Schmidt, and Adam Riess observed distant Type Ia supernovae, which led them to conclude that expansion is now accelerating~\cite{SupernovaSearchTeam:1998fmf, SupernovaCosmologyProject:1998vns, WMAP:2003elm, Huang:2005re,  Koivisto:2005mm, Fedeli:2008fh, Singh:2023bjx, Singh:2024ckh, Shaily:2024tmx, Rani:2024uah}. The inference drawn from this new acceleration was the presence of some unknown energy component with high negative pressure, known as dark energy, which is now considered to comprise around two-thirds of the total energy density of the universe~\cite{Riess:2006fw, Capozziello:2005tf, Jimenez:2003iv, Nojiri:2005pu, WMAP:2008ydk, Peebles:2002gy}. We already knew that the universe is expanding, which is an important insight that has shaped our understanding of its structure and evolution. Since then, modern cosmology has increasingly relied on astrophysical observations to tune and test theoretical models. Evidence for its expansion has come primarily from two kinds of observation: the cosmic microwave background and high-redshift Type Ia supernovae. However, from cosmological observations, it can be inferred that the Universe had two different phases in which it expanded: First, the early deceleration phase of the expansion dynamics, and then the late phase, which is characterized by accelerated expansion, which started around $ 6 $ billion years after the Big Bang and is still ongoing.

Dark energy is a kind of mysterious form of energy that is thought to pervade the universe and act as the cause of its accelerated expansion, exerting a strong negative pressure. Scientists categorize different types of dark energy based on something called the equation of state (EOS), which describes the relationship between pressure and energy density. When the EoS lies between $ -1 < w < -\frac{1}{3} $, the dark energy is referred to as quintessence type \cite{Zlatev:1998tr, Brax:1999yv, Barreiro:1999zs}. If $ w < -1 $, it's known as Phantom-type dark energy~\cite{Gorini:2002kf}. Some models allow $\omega$ to cross the critical divide $ w = -1 $ from either side; these are called Quintom-type dark energy~\cite{Feng:2004ad}. Additionally, certain cosmological theories propose an Ekpyrotic phase before the Big Bang, where the universe contracts with a very stiff equation of state, $ w \gg 1 $ \cite{Lehners:2008vx, Singh:2024tur}. Another approach to explaining the late-time acceleration of the universe is through modifications to gravity rather than introducing dark energy. These models suggest that gravity may behave differently on large scales than predicted by Einstein's general relativity, and many arise from higher-dimensional theories such as string theory and M-theory~\cite{Chiba:1999ka, Armendariz-Picon:2000nqq, Copeland:2006wr, Bamba:2012cp}. Unlike some exotic dark-energy models, these do not require negative kinetic terms. They can explain both early-time inflation and the smooth transition to late-time acceleration. Although these modified gravity theories are gaining traction as alternatives to General Relativity, they face stringent observational and theoretical constraints~\cite{Frieman:2008sn, Li:2012dt, Garriga:1999vw}.

The three key observations pillar of modern cosmology, which are the expanding universe, primordial nucleosynthesis, and the observed isotropy of cosmic microwave background radiation, are all phenomena that big-bang cosmology, based on Einstein’s field equations, was supposed to explain successfully. However, Smoot et al.~\cite{COBE:1992syq} revealed that the previous predictions of the FLRW types of models do not always fully align with our expectations. The redshift results of the extragalactic objects shown in this approach are essentially contradictory for describing the space production trend in the Big Bang-type models. And cosmic microwave background radiation was not a decisive victory for Big Bang theory either. Narlikar et al.~\cite{Narlikar:2002bk} demonstrated the possibility of a non-relativistic interpretation of cosmic microwave background radiation. All these theories have been proposed from time to time to explain such a phenomenon. In the early universe, Hoyle~\cite{Hoyle:1948zz}, Bondi and Gold~\cite{Bondi:1948qk} suggest steady state cosmology, according to which the universe sustains itself in a single persistence without beginning or end in the cosmic time scale. They have also shown that the large-scale features of the universe exhibit the same statistical properties. Furthermore, continuous creation of matter, rather than the creation of matter that occurs instantaneously, infinitely explosively at $ t = 0 $, has ensured that the mass density remains constant over time. But this formalism violated the principle of conservation of matter. To resolve this issue, Hoyle and Narlikar~\cite{hoyle1966radical} went to a field theoretic approach, namely, they introduced a massless and chargeless scalar field $ \mathcal{C} $, in the Einstein-Hilbert action to describe the matter production. There is no cosmic bang kind of singularity like that in the $ \mathcal{C} $-field theory introduced by Hoyle and Narlikar as in steady state theory~\cite{Bondi:1948qk}. 

Narlikar and Padmanabhan~\cite{Narlikar:1985llm} showed that the solutions of Einstein’s field equations allow for the presence of radiation in the form of a negative energy massless scalar creation field  $ \mathcal{C} $. Chatterjee and Banerjee ~\cite{Chatterjee:2003xs} derived the theory of Hoyle and Narlikar \cite{hoyle1966radical, hoyle1963mach, hoyle1964cfield} for a space-time of more than four dimensions, whereas Raj et al.~\cite{Raj:2007zz} studied the case for a flat FRW model with  $ \mathcal{C} $-field model and scale-dependent $ G $. More models have been constructed, such as those by Bali et al.~\cite{Bali:2009zz}, and studies of anisotropic cosmologies like the Kasner metric \cite{Kasner:1921zz} have provided important insights into the nature of singularities in general relativity. Singh et al.~\cite{Singh:2008zzu} studied several Bianchi-type models and the Kantowski-Sachs model in $ C $-field cosmology. Recently, Maurya et al. \cite{Maurya:2024wbh} analyzed various dark energy observations with the help of Hoyle–Narlikar gravity and demonstrated that such models are consistent with the current data, providing evidence for a transition from deceleration to acceleration in the expansion of the Universe.

The paper is structured as follows: In Section~\ref{sec:1}, a fundamental assessment of general relativity and various other modified theories and their cosmological implications. We present the late-time cosmic acceleration and the reasons behind this gravitational theory. In Section~\ref{sec:2}, we study the fundamental formalism of the model in the Hoyle-Narlikar Gravity and find the modified Einstein field equations. In Section~\ref{sec:3}, we present the function of the Hubble parameter $ H(z) $ and discuss the solutions of the field equations, and the behavior of the $ \mathcal{C} $-field according to our model. In Section~\ref{sec:4}, we constrain the Hubble tension and the density parameters using recent observational data, and also perform a statistical model comparison with the $ \Lambda $CDM model employing the Akaike and Bayesian information criteria. In Section~\ref{sec:5}, we concentrate on the physical interpretation and diagnostic profile of the model. In Section~\ref{sec:6}, we perform the thawing and freezing analysis in the $ w-w'$ plane to distinguish between different classes of dark energy models. Finally, we summarize the results and the consequences of the model in the concluding section~\ref{sec:7}.

\section {Hoyle-Narlikar Creation Field Theory}\label{sec:2}

\qquad The Hilbert-Einstein action with a creation matter field $ \mathcal{C} $ is the basic action functional for the Hoyle-Narlikar theory of gravity, which was introduced in 1964 by Fred Hoyle and Jayant V. Narlikar. It modifies standard General Relativity to allow for the continuous creation of matter, and favours the Steady State model of the universe.

The standard Einstein-Hilbert action includes the gravitational term, the matter term, and the specific $ \mathcal{C} $-field term, and is expressed as
\begin{equation}\label{eq:1a}
  S=\int \sqrt{-g}\left[\frac{R}{16\pi G}+\mathcal{L}_{m}+\mathcal{L}_{C}\right]d^{4}x,
\end{equation}
where $ R $ is The Ricci scalar indicating the spacetime curvature, $ \mathcal{L}_{m} $ is the Lagrangian density for ordinary matter, and  $ \mathcal{L}_{C} $ is the Lagrangian density for the creation field ($ \mathcal{C} $-field). 

The $ \mathcal{C} $-Field Lagrangian is defined as $ \mathcal{L}_{C}=-\frac{1}{2}\zeta \mathcal{C}_{i}\mathcal{C}^{i} $, where $ \zeta  $ is a coupling constant ($ \zeta >0 $) and $ \mathcal{C}_{i}=\partial _{i}\mathcal{C} $ represents the field's gradient. The $ \mathcal{C} $-field is a massless scalar field of negative energy, and acts as a reservoir of negative energy. As the universe expands, the $ \mathcal{L}_{C} $-field creates new matter (positive energy) out of empty space, which preserves a constant mass density and avoids the necessity for a Big Bang singularity.

The contribution of the $ \mathcal{L}_{C} $-field to the field equations is represented by its energy-momentum tensor $  T_{ij}^{(\mathcal{C})} $, which is given by  
\begin{equation}\label{eq:1b}
    T_{ij}^{(\mathcal{C})} = -s\zeta\left( \mathcal{C}_{i} \mathcal{C}_{j}-\frac{1}{2}g_{ij} \mathcal{C}^{\alpha } \mathcal{C}_{\alpha }\right),
\end{equation}
where we have introduced a new parameter $ s = \pm 1 $ to keep track of the sign convention. The negative sign in this tensor indicates its repulsive gravitational nature, which drives the universe's expansion.
By varying this action with respect to the metric $ g_{ij} $ and the $ \mathcal{C} $-field, we obtain modified Einstein field equations, which describe how the spacetime curvature (gravity) and matter creation ($ \mathcal{C} $-field) evolve together, forming the basis of the Hoyle-Narlikar theory \cite{hoyle1966radical, hoyle1963mach, hoyle1964cfield, narlikar1973singularity, Narlikar:1985llm, Narlikar:2002bk}.
\begin{equation}\label{eq:1}
    R_{ij} - \frac{1}{2} R g_{ij} = -8\pi \left[ T_{ij}^{(m)} + T_{ij}^{(\mathcal{C})} \right],
\end{equation}
where the energy momentum tensors $ T_{ij}^{(m)} $, $ T_{ij}^{(\mathcal{C})} $, corresponding to the perfect fluid matter source and the creation field, respectively, are given as
\begin{equation}\label{eq:2}
    T_{ij}^{(m)} = (\rho + p) u_{i} u_{j} + p g_{ij},
\end{equation}
and from Eq. (\ref{eq:1b}), we get
\begin{equation}\label{eq:3}
    T_{ij}^{(\mathcal{C})} = -s \zeta \left( \mathcal{C}_{i} \mathcal{C}_{j} - \frac{1}{2} \mathcal{C}^2 g_{ij} \right),
\end{equation}
where $ \rho $, $ p $ are the energy density and the matter pressure, respectively. In co-moving coordinate system, we consider the four-velocity vector $ u^{i}=(0,0,0,-1) $ such that $ u_{i} u^{i} =-1 $ and the creation field $ \mathcal{C}\propto t $. 

To analyze cosmological dynamics, we consider the flat FLRW metric in the form \cite{Ryden:1970vsj}
\begin{equation}\label{eq:4}
     ds^2 = -dt^2 + a^2dx_i^2, ~i=1,2,3,
\end{equation}
where $ a(t) $ is the scale factor that characterizes the expansion of the Universe over time.

Now, by substituting Eqs.~(\ref{eq:2}), (\ref{eq:3}), and (\ref{eq:4}) into Eq.~(\ref{eq:1}), we obtain the modified field equations as follows
\begin{equation}\label{eq:5}
   8\pi\rho= 3H^2 + 4 \pi s \zeta \dot{\mathcal{C}}^2,
\end{equation}
\begin{equation}\label{eq:6}
  8\pi p =  -\left(2\dot{H} + 3H^2 \right)+ 4 \pi s \zeta \dot{\mathcal{C}}^2.
\end{equation}

The energy conservation equation is given by
\begin{equation}\label{eq:7}
    \dot{\rho} - s\zeta \dot{\mathcal{C}}\ddot{\mathcal{C}} + 3H(\rho + p - s \zeta \dot{\mathcal{C}}^2 ) = 0,
\end{equation}
or 
\begin{equation}\label{eq:7a}
  \dot{\rho} + 3H(\rho + p) = s \dot{\mathcal{C}} \left( \ddot{\mathcal{C}} + 3H \dot{\mathcal{C}} \right).
\end{equation}

The equation (\ref{eq:7a}) governs the creation field as
\begin{equation*}
\ddot{\mathcal{C}} + 3H \dot{\mathcal{C}} = 0 .
\end{equation*}
which ensures the internal consistency of the modified gravitational dynamics and describes the evolution of the creation field in an expanding FLRW background. We then adopt the ansatz in which the creation field $ \mathcal{C} $ varies with cosmic time $ t $ as
\begin{equation}\label{eq:cf}
 \mathcal{C} = k \tanh(\alpha t),
\end{equation}
where $ k $ sets the late-time value and $ \alpha $ controls the characteristic time scale. In our analysis, we adopt $ k=0.16 $ and $ \alpha=1.05 $. 

The creation field $ \mathcal{C} $ empowers continuous matter creation without violating energy conservation and produces a repulsive gravitational effect in the direction of the cosmological constant $ \Lambda$, which supports both early and late-time accelerated expansion of the Universe. The case $ \zeta = 0 $ corresponds to the model in the GR form, and $ \zeta > 0 $ describes the interaction of the creation field and matter model.

\begin{figure}
    \centering
    \includegraphics[width=0.5\linewidth]{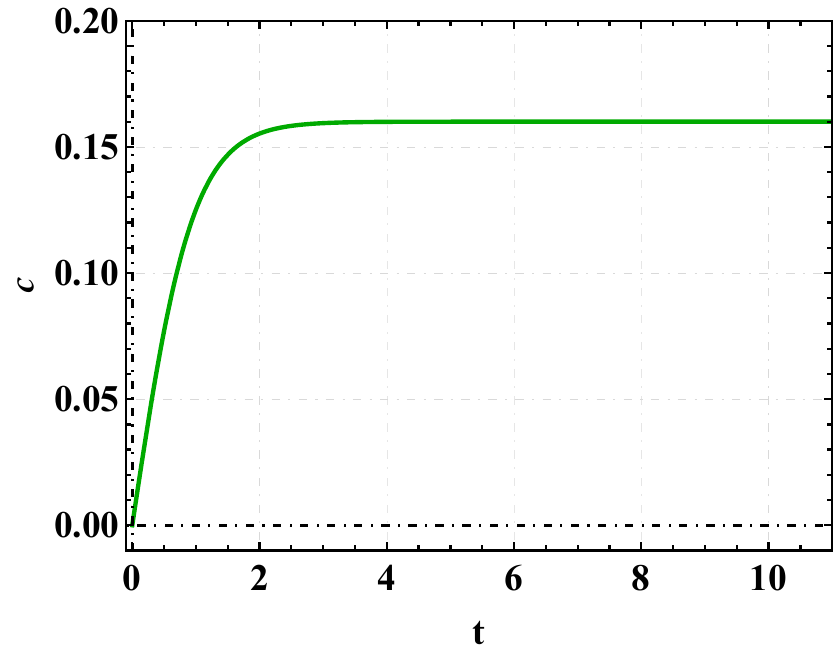}
    \caption{ The evolution of the creation field $ \mathcal{C} $ with the cosmic time $ t $ for the values $ k = 0.16 $ and $\alpha = 1.05$.}
    \label{fig:1}
\end{figure}

The creation field evolves with cosmic time according to Eq. (\ref{eq:cf}) since homogeneity and isotropy of the FLRW background rule out any spatial dependence. This functional form is smooth, bounded, and monotonic, ensuring a regular time evolution of the creation field. It provides a simple representation in which the field grows gradually at early times and asymptotically approaches a constant value at late times, thereby capturing the essential physics of a uniform matter--creation process. $ \zeta > 0 $ as a coupling constant between matter and the creation field.

For $ s = +1 $, the creation field contributes with a negative effective energy density, while for $ s = -1 $, the sign is reversed and the creation field contributes positively. The precise condition on the sign of $ \zeta $ will be obtained later when we apply the present-day boundary condition $ H(a_0)=H_0 $ and introduce the matter density 
parameter $ \Omega_{m0} = 8\pi \rho_0 / (3H_0^2) $. In this work, we adopt the convention $ s = -1 $, which leads to a consistent formulation of the model and allows a positive value of $ \zeta $.

In that case, the component $ T_{44}^{(\mathcal{C})} = -\frac{s}{2} \zeta \mathcal{\dot{C}}^2 $ determines the effective contribution of the creation field. Depending on the value of $s$, the creation field can act as a repulsive source term for $ s = +1 $ or as a positive-energy contribution for $ s = -1 $.

As shown in Fig.~\ref{fig:1}, the field starts from zero at $ t = 0 $, grows rapidly during the early epoch, and asymptotically approaches the constant value $ k $ at late times. This behaviour provides a smooth and bounded evolution for the creation field. It rises during the early stage of the universe and settles to a constant value at late times, which matches how the $ \mathcal{C} $-field is expected to behave in Hoyle-Narlikar gravity. Other choices, such as exponential or power laws, can grow without limit or lead to unrealistic behaviour, so the $ \tanh $ form is a practical and well-behaved option.

\section{Parametrization of Hubble parameter}\label{sec:3}

In this paper, we employ the cosmological parameterization method to solve the field equations and examine the dynamical evolution of the universe. Various studies have proposed different parameterizations of cosmological quantities to characterize significant epochs, particularly the transition from early-time deceleration to late-time acceleration. The parameters introduced through this approach can be constrained using observational datasets such as Hubble data, Pantheon$^{+}$, and their combination with BAO measurements. The literature presents a range of parameterization schemes, commonly based on the deceleration parameter ($q$), the Hubble parameter ($H$), or the equation of state (EoS) parameter. In our analysis, we adopt the following specific parameterization of the Hubble parameter
\begin{equation}\label{eq:8}
    H(z) = H_0 \sqrt{(1 + z)^{3} \Omega_m + (1 + z)^{4} \Omega_{r} + \Omega_{\Lambda}},
\end{equation}
where $ H_0 $, $ \Omega_m $, $ \Omega_{r} $, and $ \Omega_{\Lambda} $ represent the present value of the Hubble parameter, density parameter for matter, density parameter for radiation, and density parameter for dark energy, respectively. These density parameters are defined by $\Omega_i = \rho_{i0}/\rho_c$; $\rho_c = 3H_0^2/(8\pi G)$, where $\rho_{i0}$ denotes the present energy density of the component ($ i = m,r, \Lambda$). 

In this work, we use the $ \Lambda $CDM form of $ H(z) $ for the background expansion. The Hoyle-Narlikar field equations with a creation field do not yield a simple analytic expression for $ H(z) $ without making extra assumptions, so adopting the standard $ \Lambda $CDM form keeps the analysis transparent. This also allows us to focus mainly on how the $ \mathcal{C} $-field influences the cosmological dynamics without changing the well-tested expansion history.

Substituting the relation $ H= \frac{\dot{a}}{a} $ and $  a= \frac{1}{1+z} $ into Eq.~(\ref{eq:8}), we obtain cosmic time $ t $ as a function of redshift as
\begin{equation}\label{eq:10}
    t = \frac{-2 \Omega_m + 6 \Omega_{r} (1 + z) - \Omega_{\Lambda}}{12 H_0 \Omega_{r}^{3/2} (1 + z)^3}.
\end{equation}
It links cosmic time $ t $ with the observable redshift $ z $, providing a useful connection between the theoretical formulation and cosmological data.

\section{Statistical observation for model parameters} \label{sec:4}

Emphasizing the importance of a comprehensive assessment of parameter values when investigating cosmological models, the present section performs observational analyses of the prevailing scenario~\cite{DESI:2024mwx}. We find the best-fit values of the parameters using Markov chain Monte Carlo (MCMC) analysis with \texttt{emcee} as the Python implementation~\cite{Foreman-Mackey:2012any, Singh:2024tur, Shaily:2026sha}. We impose the following prior ranges for the parameters: $ H_0 \in [60, 80] $, $ \Omega_m \in [0.2, 1] $, $ \Omega_{r} \in [0, 0.2] $, and $ \Omega_{\Lambda} \in [0.5, 1] $.

Our analysis combines data from the $ H(z) $ and Pantheon datasets, and the form of the $ \chi^2 $ function for various datasets is provided below.

\begin{figure}
   \centering
    \subfloat[]{\label{fig:2a}\includegraphics[width=0.5\linewidth]{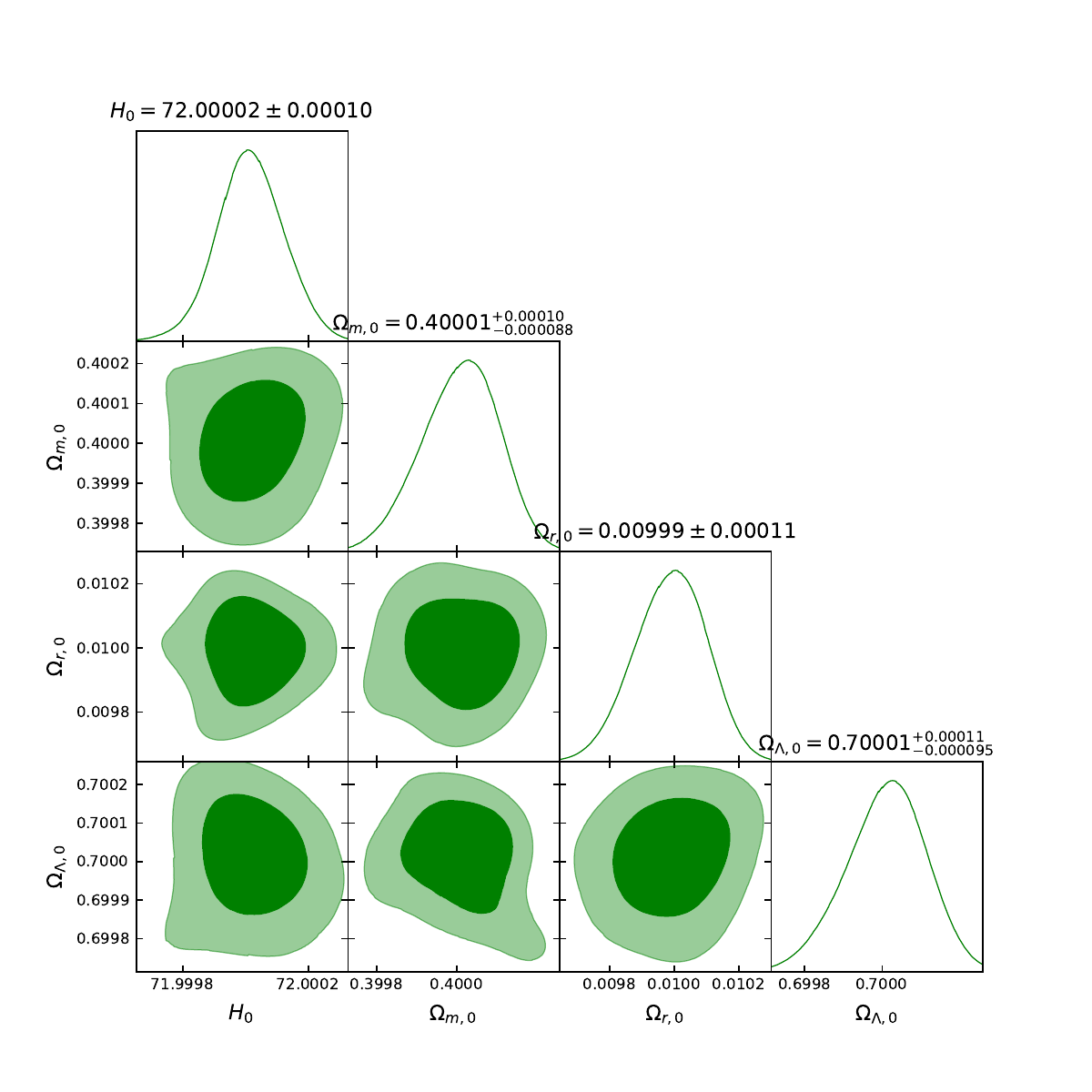}}\hfill
    \subfloat[]{\label{fig:2a}\includegraphics[width=0.5\linewidth]{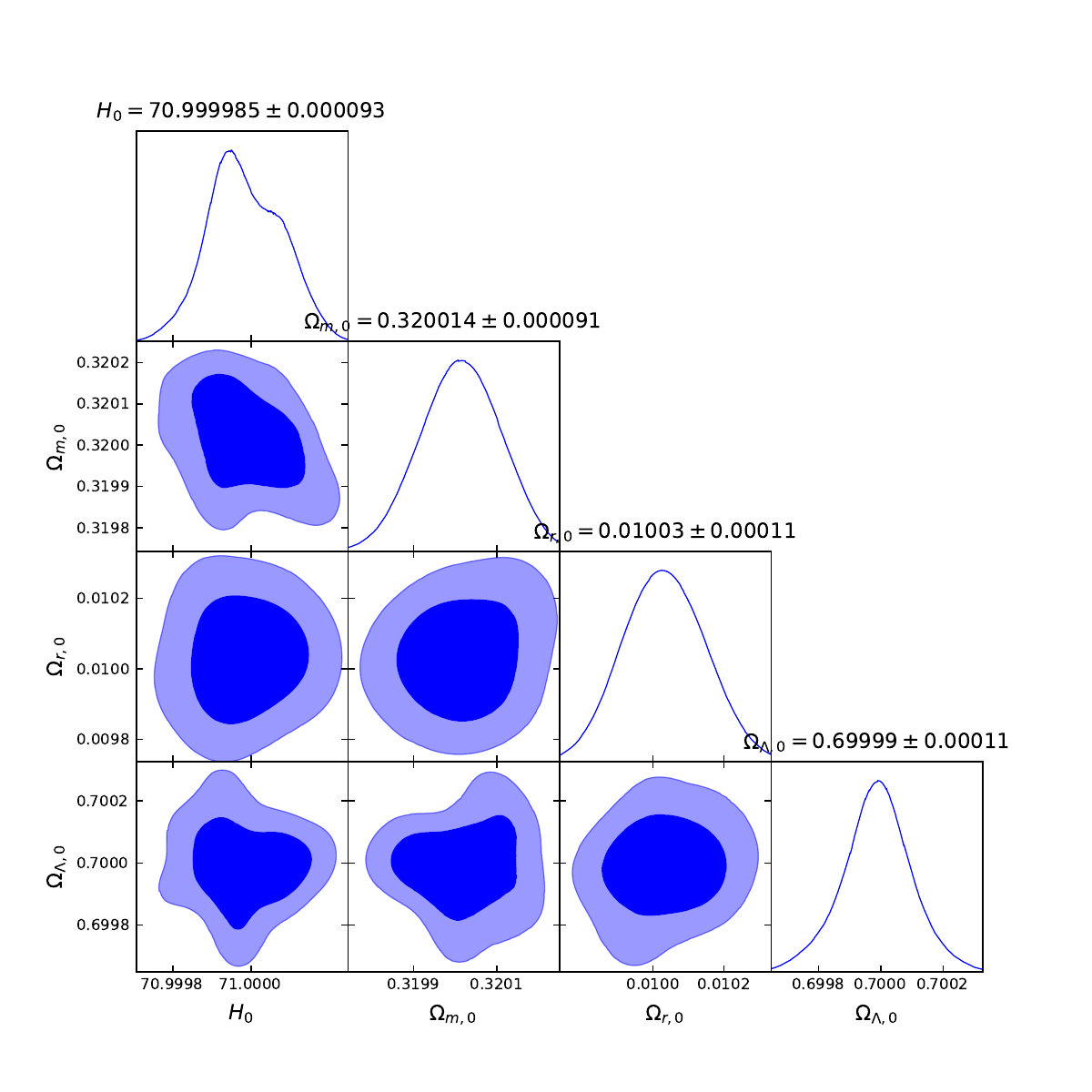}}\par
     \subfloat[]{\label{fig:2a}\includegraphics[width=0.5\linewidth]{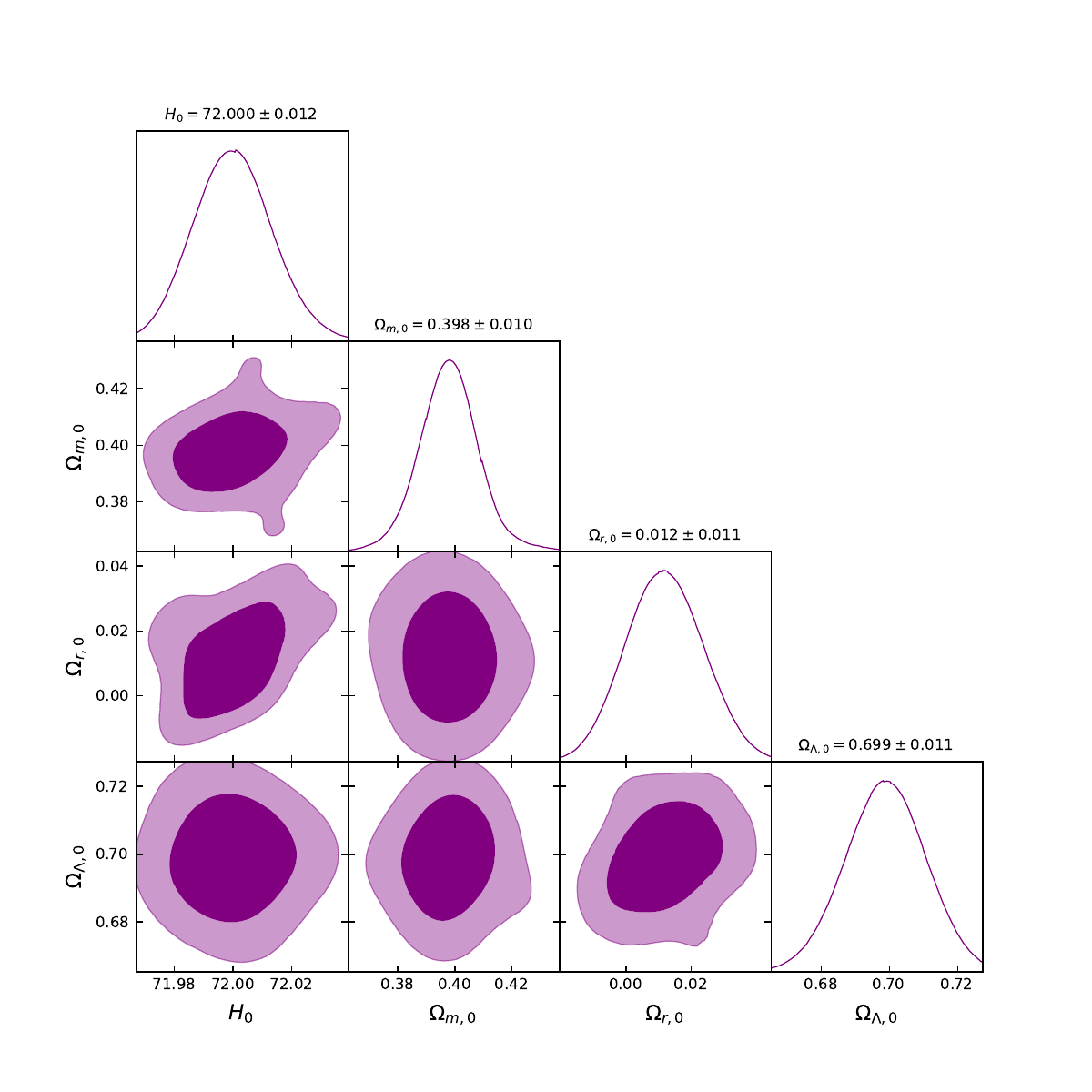}}\par
    \caption{ Posterior distributions for the model parameter pairs with the iso-surfaces correspond to 1$\sigma$ and 2$\sigma$ CLs for all data compilations.}    
     \label{Fig:2}
\end{figure}

\subsection{Hubble Dataset}

The Hubble parameter $ H(z) $ is a measure of the rate of expansion of the universe at a redshift $ z $. We can estimate using the relationship
\begin{equation}\label{eq:14}
H(z) = \frac{\dot{a}}{a} = -\frac{1}{1+z} \frac{dz}{dt} \simeq -\frac{1}{1+z} \frac{\Delta z}{\Delta t},
\end{equation}
where $ \dot{a} $ is a derivative of the scale factor $ a(t) $ with respect to time. Here $ \Delta z $ and $ \Delta t $ are the change in redshift and cosmic time of two passively evolving galaxies. The spectroscopic shift $ \Delta z $ is determined by a set of spectroscopy, and $ \Delta t $ is derived by the differential age method. All this together makes it possible to observationally measure without any model for $ H(z) $.
In this work, we use a compilation of 77 values of the Hubble parameter covering the interval $ 0 \leq z \leq 2.360 $~\cite{Singh:2024kez}. To test the theoretical predictions against the observations, we use the chi-squared statistic.
\begin{equation}\label{eq:15}
\chi^2_{\text{HD}} = \sum_{i=1}^{77} \left( \frac{H_{\text{th}}(z_i) - H_{\text{obs}}(z_i)}{\sigma_H(z_i)} \right)^2,
\end{equation}
where $ H_{\text{obs}}(z_i) $ and $ H_{\text{th}}(z_i) $ are the observed and theoretical Hubble parameter at redshift $ z_i $, respectively, and $ \sigma_H(z_i) $ is the observational uncertainty.

\subsection{Pantheon$^{+}$ Dataset}

The expansive Pantheon$^{+}$ + SH0ES dataset consists of $ 1701 $ Type Ia Supernovae (SNIa) obtained from various surveys~\cite{Brout:2022vxf}. This dataset covers the redshift range ($ 0.001 < z < 2.2613 $ ), enabling a detailed study of cosmic expansion. Building on previous compilations like Union~\cite{SupernovaCosmologyProject:2008ojh}, Union 2~\cite{Amanullah:2010vv}, Union 2.1~\cite{SupernovaCosmologyProject:2011ycw}, JLA~\cite{SDSS:2014iwm}, and Pantheon~\cite{Pan-STARRS1:2017jku}, Pantheon$^{+}$~\cite{Scolnic:2021amr} includes the latest observations, enhancing its value for cosmological research through precise distance measurements using SNIa as standard candles.  The Pantheon$^{+}$ + SH0ES dataset refines systematic uncertainties, leading to more precise estimates of the cosmological parameters. Its primary purpose is to compare actual distance measures with theoretical expectations, particularly in the context of the Hubble tension and dark energy studies.
 
\begin{equation}\label{eq:16}
    \chi^2_{\text{SNe}} = \mathbf{D}^{T} \mathbf{C}^{-1}_{\text{SNe}} \mathbf{D},
\end{equation}
Here, $ \mathbf{C}_{\text{SNe}} $ \cite{Brout:2022vxf} denotes the covariance matrix for the Pantheon$^{+}$ + SH0ES dataset, capturing both statistical and systematic uncertainties. The vector $ \mathbf{D} $ is defined as
\begin{equation}\label{eq:17}
    \mathbf{D} = m_{\text{B}i} - M - \mu^\text{th}(z_i),
\end{equation}
where $ m_{\text{B}i} $ is the apparent magnitude and $M$ is the absolute magnitude. In addition, $ \mu^\text{th}(z_i) $ represents the distance modulus of the theoretical model, which is calculated using the equation below.
\begin{equation}\label{eq:18}
    \mu^\text{th}(z_i) = 5\log_{10} \left( \frac{D_L(z_i)}{1 \text{ Mpc}} \right) + 25,
\end{equation}
In this expression, $ D_L(z) $ represents the luminosity distance, defined as the distance to an object based on its luminosity and the observed brightness of the light it emits. For the specified model, it can be calculated as
\begin{equation}\label{eq:19}
    D_L(z) = c(1+z) \int_0^z \frac{dz'}{H(z', \bm{\theta})},
\end{equation}
where $ \bm{\theta} = (H_0, \Omega_m, \Omega_{r},\Omega_{\Lambda}) $. In contrast to the Pantheon dataset, the Pantheon$^{+}$ + SH0ES compilation successfully reduces the degeneracy between $ H_0 $ and $ M $ by reformulating the vector $ \mathbf{D} $ as
\begin{equation}\label{eq:20}
    \hat{\mathbf{D}} = \begin{cases}
        m_{\text{B}i} - M - \mu^\text{Ceph}_i, & i \in \text{Cepheid hosts}, \\
        m_{\text{B}i} - M - \mu^\text{th}(z_i), & \text{otherwise}.
    \end{cases}
\end{equation}
where $ \mu_i^{\text{Ceph}} $ is independently estimated using Cepheid calibrators. Consequently, Eq.~(\ref{eq:16}) becomes $ \chi^2_{\text{SNe}} = \tilde{D}^T C_{\text{SNe}}^{-1} \tilde{D} $.

\begin{figure}
   \centering
    \subfloat[]{\label{fig:3a}\includegraphics[width=0.46\linewidth, height=0.35\linewidth]{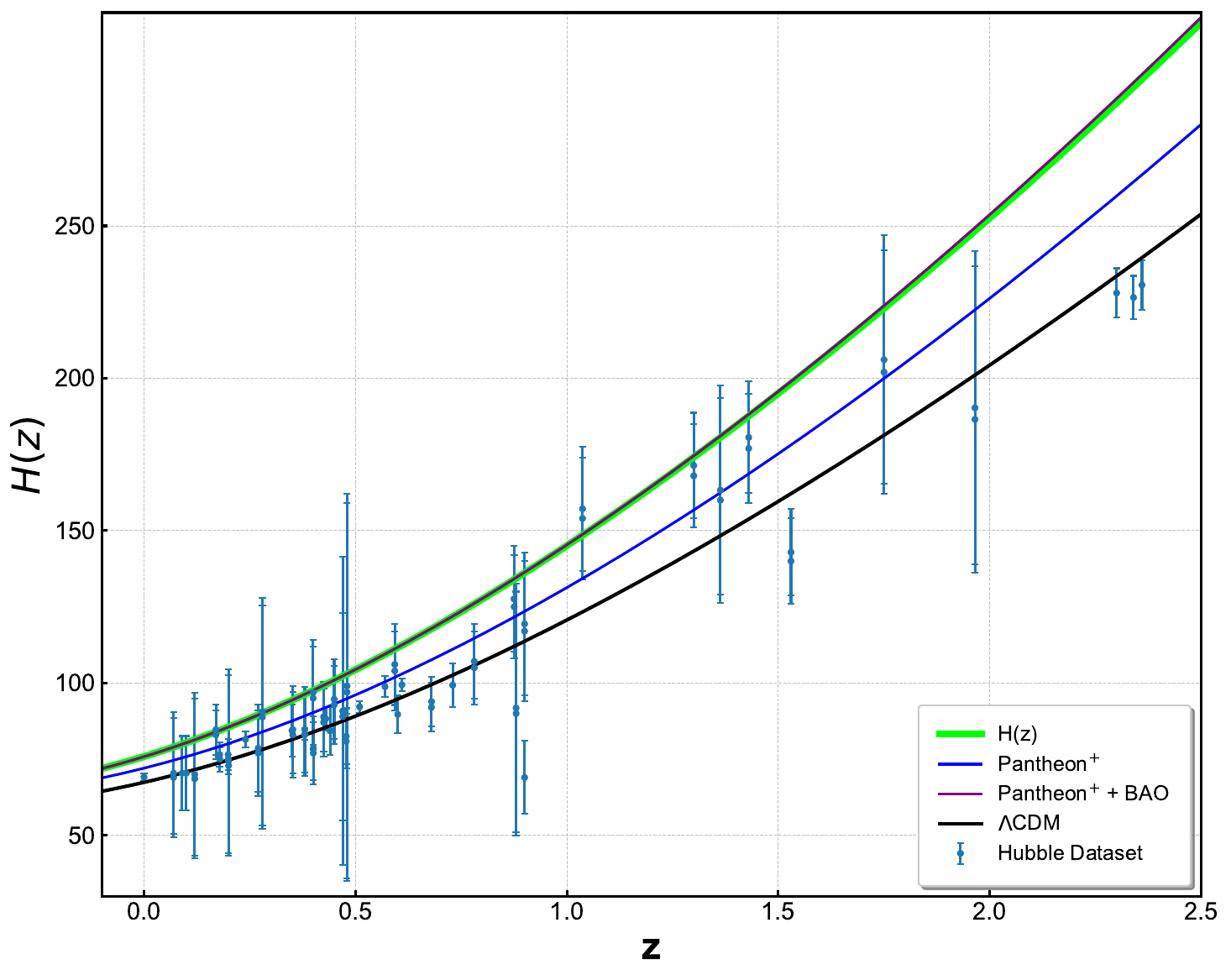}}\hfill
    \subfloat[]{\label{fig:3b}\includegraphics[width=0.46\linewidth, height=0.35\linewidth]{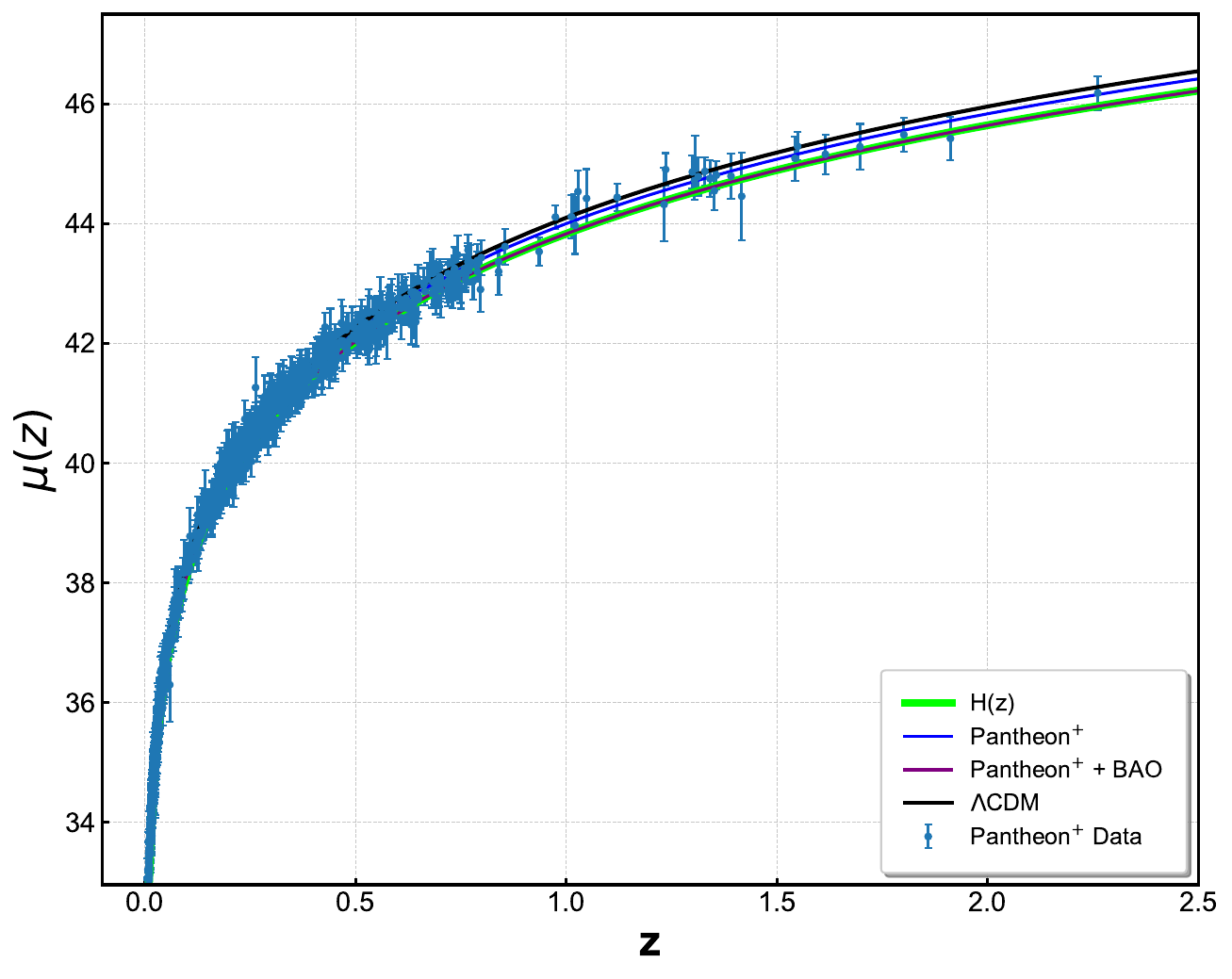}}
    \caption{The Error Bar plots exhibit the consistency of our model with the $ \Lambda $CDM model across all data compilations.}
    \label{Fig:3}
\end{figure}

\subsection{BAO Dataset}

Since BAO observations provided efficient ultimate constraints in the analysis presented in this work, we include a description of the BAO studies. BAO studies reveal oscillations that were excited in the early Universe, originating from cosmic perturbations, within a fluid comprising photons, baryons, and dark matter that are loosely coupled through Thomson scattering. BAO observations include the Sloan Digital Sky Survey (SDSS), the Six Degree Field Galaxy Survey (6dFGS), and the Baryon Oscillation Spectroscopy Survey (BOSS). The dilation scale, $ D_v(z) $, is expressed as follows~\cite{SDSS:2005xqv,Blake:2011en, SDSS:2009ocz}

\begin{equation}\label{eq:21}
D_v(z) = \left( \frac{d_A^2(z) z}{H(z)} \right)^{1/3},
\end{equation}
Here, the comoving angular diameter distance, $ d_A(z) $, is defined by
\begin{equation}\label{eq:22}
d_A(z) = \int_0^z \frac{dz'}{H(z')}.
\end{equation}
The chi-squared function used in the BAO analysis is given by
\begin{equation}\label{eq:23}
\chi^2_{BAO} = A^T C^{-1}_{BAO} A.
\end{equation}
In this expression, $ A $ is specific to the survey being analyzed, and $ C^{-1} $ represents the inverse of the covariance matrix~\cite{Giostri:2012ek}.

\subsection{The $ \Lambda $CDM Model}

The most well-known and straightforward model of dark energy in standard cosmology is the $ \Lambda $CDM model, where $ \Lambda $ denotes the cosmological constant. In this case, the energy density and pressure of dark energy are time-independent. They are represented by $ \rho_{\Lambda} \equiv \frac{\Lambda}{8\pi G} $ and $ p_{\Lambda} = -\rho_{\Lambda} $, respectively. This corresponds to a constant equation of state (EoS) parameter $ w_\Lambda = -1 $.

The Hubble parameter in a spatially flat Friedmann–Lemaître–Robertson–Walker (FLRW) universe in the framework of the $ \Lambda $CDM is given by
\begin{equation}\label{eq:24}
H(z) = H_0 \left[\Omega_m(1 + z)^3 + (1 - \Omega_m)(1 + z)^{3(1 + w)}\right]^{1/2},
\end{equation}
where $ H_0 $ and $ \Omega_m $ are the current value of the Hubble parameter and the density parameter of matter.

Table~\ref{tabparm} presents the constrained values of the model parameters for all three datasets. The constrained values are evaluated by the Posterior distributions for the model parameter pairs, with the iso-surfaces corresponding to 1$\sigma$ and 2$\sigma$ CLs for all data compilations shown in Fig.~\ref{Fig:2}. The Error Bar plots exhibit the consistency of our model with $ \Lambda $CDM for all data compilations shown in Fig.~\ref{Fig:3}. The Hubble parameter $ H> 0 $ in the evolution indicates that our model is in an expanding state of the Universe.

 \begin{table}
 \caption{Results of the data analysis for this model}
 \begin{center}
 \label{tabparm}
 \resizebox{\textwidth}{!}{%
 \begin{tabular}{l c c c c c c} 
 \hline\hline
 \\ 
 {Dataset} \,\,\,\,\,  &  \,\, \,   $H_0$\footnote{$km s^{-1} Mpc^{-1}$}   \,\,\,\,\,  &  \,\, \,  $\Omega_m $\footnote{$\Omega_i \equiv \Omega_{i,0}$ $(i = m, r, \Lambda)$} \,\,\,\,\,  &  \,\, \,   $\Omega_{r}$   \,\,\,\,\,  &  \,\, \,  $\Omega_{\Lambda}$ \,\,\,\,\,  &  \,\, \,  $ z_{tr} $\footnote{Transitions from deceleration to acceleration} \,\,\,\,\,  &  \,\, \, $ q_0 $\footnote{The present value of the $ q $}
 \\ 
 \\
 \hline 
 \\
 {$H(z)$}   \,\,\,\,\,  &  \,\, \,   $72.00 \pm 0.00010$   \,\,\,\,\,  &  \,\, \,  $0.40^{+0.00010}_{-0.000088} $ \,\,\,\,\,  &  \,\, \,   $0.009 \pm 0.00011$   \,\,\,\,\,  &  \,\, \,  $0.70^{+0.00011}_{-0.000095}$   \,\,\,\,\,  &  \,\, \,   $0.482$   \,\,\,\,\,  &  \,\, \,   $-0.441$
 \\
 \\
 {Pantheon$^{+}$ }   \,\,\,\,\,  &  \,\, \,   $70.99 \pm 0.000093$   \,\,\,\,\,  &  \,\, \,  $0.30\pm 0.000091 $ \,\,\,\,\,  &  \,\, \,   $0.01 \pm 0.00011$   \,\,\,\,\,  &  \,\, \,  $0.69 \pm 0.00011$   \,\,\,\,\,  &  \,\, \,   $0.584$   \,\,\,\,\,  &  \,\, \,   $-0.514$
 \\
 \\
 {Pantheon$^{+}$ +  BAO}   \,\,\,\,\,  &  \,\, \,   $72.00 \pm 0.012$   \,\,\,\,\,  &  \,\, \,  $0.398 \pm 0.010 $ \,\,\,\,\,  &  \,\, \,   $0.012 \pm 0.011$   \,\,\,\,\,  &  \,\, \,  $0.69 \pm 0.011$   \,\,\,\,\,  &  \,\, \,   $0.477$   \,\,\,\,\,  &  \,\, \,   $-0.440$
 \\
 \\ 
 \hline\hline  
 \end{tabular}%
 }
 \end{center}
 \label{Table 1}
 \end{table}

\begin{figure}
    \centering
     \subfloat[]{\label{fig:cov}\includegraphics[width=0.46\linewidth, height=0.35\linewidth]{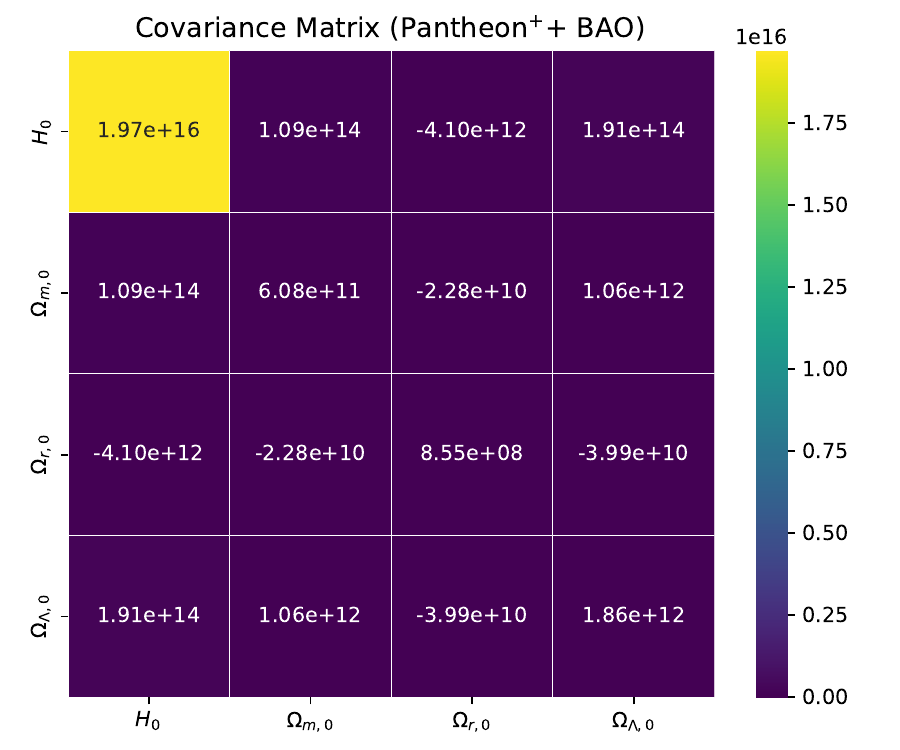}}\hfill
    \subfloat[]{\label{fig:cor}\includegraphics[width=0.46\linewidth, height=0.35\linewidth]{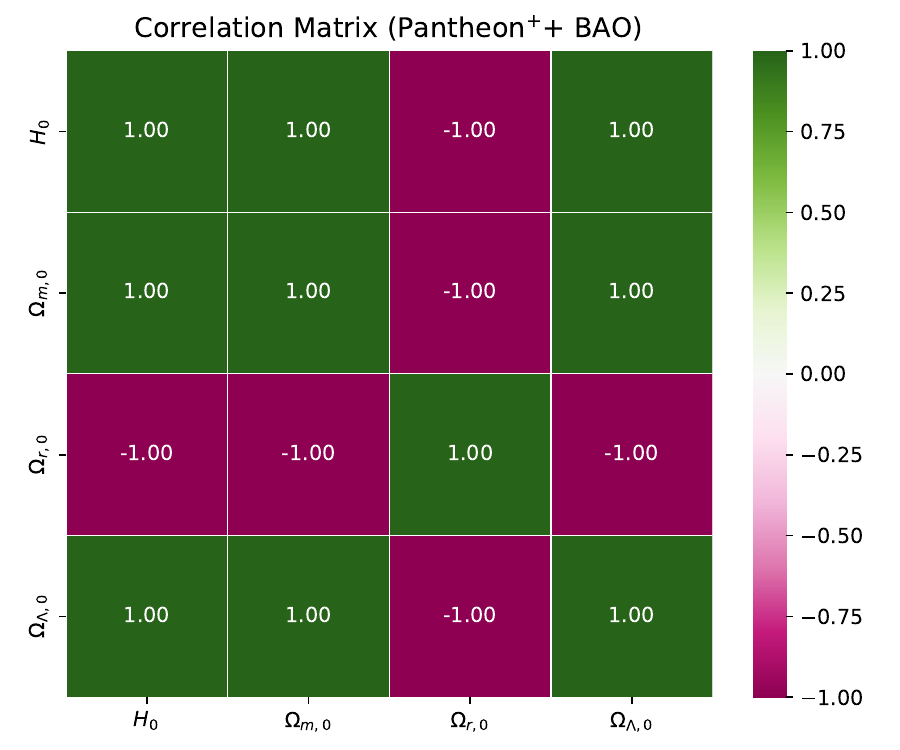}}
\caption{Covariance and correlation matrices among the model parameters for all data compilations.}
\label{Fig:corr}
\end{figure}

The covariance matrix in Fig.~\ref{fig:cov} is obtained from the joint analysis of the Pantheon$^+$ and BAO datasets. It reflects the uncertainties and correlations between the cosmological parameters $ H_0 $, $ \Omega_{m,0} $, $ \Omega_{r,0} $, and $ \Omega_{\Lambda,0} $. The diagonal elements represent the variances of each parameter, indicating their associated uncertainty. The terms except diagonal terms in the matrix indicate covariances, where positive values imply that two parameters increase or decrease together, while negative values suggest inverse relationships. The model parameters $ H_0 $, $ \Omega_{m,0} $, and $ \Omega_{\Lambda,0} $ are positively correlated, while $ H_0 $ and $ \Omega_{r,0} $ are negatively correlated. The intensity of the color in the matrix represents the magnitude of the covariance values, with the brightest region indicating the highest variance for $ H_0 $.

The correlation matrix in Fig.~\ref{fig:cor} is derived from the combined Pantheon$^+$ and BAO dataset. It shows the correlations among the model parameters as  $ H_0 $, $ \Omega_{m,0} $, $ \Omega_{r,0} $, and $ \Omega_{\Lambda,0} $. The diagonal elements are unity, indicating perfect self-correlation, while the off-diagonal elements are significantly large, reflecting strong inter-parameter correlations. Specifically, $ H_0 $, $ \Omega_{m,0} $, and $ \Omega_{\Lambda,0} $ are positively correlated, whereas $ \Omega_{r,0} $ shows strong negative correlation with the others. This indicates a high degree of parameter degeneracy, where variation in one parameter can significantly influence the others.

\subsection{Information Criteria}

\qquad To evaluate the reliability of our MCMC constraints, we employ two standard statistical tools: 
The Akaike Information Criterion ($\mathrm{AIC}$) and the Bayesian Information Criterion ($\mathrm{BIC}$)~\cite{Akaike:1974vps, Liddle:2007fy, Tan:2011pa, Vrieze:2012, Arevalo:2016epc, Schwarz:1978tpv}. 
The AIC is defined as
\begin{equation}\label{24a}
    \mathrm{AIC} = \chi^{2}_{\min} + 2\kappa,
\end{equation}
where $ \kappa $ represents the number of free parameters in the model. To compare our results with the standard $ \Lambda $CDM framework, we compute $\Delta \mathrm{AIC} = \left|\, \mathrm{AIC}_{\mathrm{Hoyle–Narlikar}} - \mathrm{AIC}_{\Lambda\mathrm{CDM}} \,\right|.$
A value $\Delta\mathrm{AIC} < 2 $ indicates strong support for the Hoyle–Narlikar model, while $ 4 < \Delta\mathrm{AIC} \leq 7 $ corresponds to moderate support. If $ \Delta\mathrm{AIC} > 10 $, the observational data do not provide significant evidence in favor of the Hoyle–Narlikar gravity model~\cite{Jeffreys:1939xee, Solanki:2024yfr}. Similarly, the BIC is given by
\begin{equation}\label{24b}
    \mathrm{BIC} = \chi^{2}_{\min} + \kappa \ln(N),
\end{equation}
where $N$ denotes the total number of observational points used in the MCMC analysis. The interpretation follows the usual criteria: $\Delta\mathrm{BIC} < 2$ suggests strong support for the Hoyle–Narlikar model, $2 \leq \Delta\mathrm{BIC} < 6$ implies moderate support, and $\Delta\mathrm{BIC} > 6$ indicates weak or negligible evidence.

The computed $\mathrm{AIC}$ and $\mathrm{BIC}$ values for the Hoyle–Narlikar gravity model and the $\Lambda$CDM model are summarized in Table~\ref{table2}. The $\mathrm{AIC}$ results indicate no significant preference between the two models for the $H(z)$ dataset, while Pantheon$^{+}$ and the joint datasets favor $\Lambda$CDM. In contrast, the $\mathrm{BIC}$ results show moderate support for $\Lambda$CDM in the Hubble dataset and strong evidence in favor of $\Lambda$CDM for the Pantheon$^{+}$ and combined datasets.

\begin{table*}[t]
\centering
\caption{The summary of the information criteria of the model, and the $ \Lambda $CDM.}
\renewcommand{\arraystretch}{1.3}
\setlength{\tabcolsep}{6pt}
\begin{tabular}{lcccccc}
\toprule
\hline
Parameters & $\chi^2$ & $\Delta \chi^2$  & $ \mathrm{AIC} $ & $ \Delta \mathrm{AIC} $ & $ \mathrm{BIC} $  & $ \Delta \mathrm{BIC} $   \\
\hline
\midrule

H(z)  &
$130.507$ &
$3.435$ &
$138.507$ &
$0.565$ &
$147.882$ &
$5.253.$ \\
 $ \Lambda $CDM
&
\textcolor{bestfitblue}{133.942} &
\textcolor{bestfitblue}{-} &
\textcolor{bestfitblue}{ 137.942} &
\textcolor{bestfitblue}{-} &
\textcolor{bestfitblue}{142.629} &
\textcolor{bestfitblue}{-} \\
\hline
\addlinespace

{Pantheon$^{+}$} &
$812.518$ &
$0$ &
$820.518$ &
$ 4$ &
$842.272$ &
$14.877$ \\
$ \Lambda $CDM
&
\textcolor{bestfitblue}{812.518} &
\textcolor{bestfitblue}{-} &
\textcolor{bestfitblue}{816.518} &
\textcolor{bestfitblue}{-} &
\textcolor{bestfitblue}{827.395} &
\textcolor{bestfitblue}{-} \\
\hline
\addlinespace

{Pantheon$^{+}$ +  BAO} & 
$1081.611$ &
$44.074$ &
$1089.611$ &
$48.074$ &
$1111.588$ &
$59.078$ \\
$ \Lambda $CDM
&
\textcolor{bestfitblue}{1037.537} &
\textcolor{bestfitblue}{-} &
\textcolor{bestfitblue}{1041.537} &
\textcolor{bestfitblue}{-} &
\textcolor{bestfitblue}{1052.510} &
\textcolor{bestfitblue}{-} \\
\hline
\midrule
\end{tabular}
\label{table2}
\end{table*}

\section{COSMOGRAPHIC ANALYSIS OF THE MODEL}\label{sec:5}

In this section, we characterize the significance of some important physical quantities such as the deceleration parameter, the jerk parameter, the statefinder parameters, and the energy conditions of the cosmological models.

The deceleration parameter is defined as
\begin{equation}\label{eq:25}
  q = -\frac{1}{H^2} \frac{\ddot{a}}{a} = -\left(1 + \frac{\dot{H}}{H^2} \right),
\end{equation}
which can be reduced in terms of redshift $z$ as
\begin{equation}\label{eq:26}
 q = -1 + \frac{(1 + z) \left(3 \Omega_m (1 + z)^2 + 4 \Omega_{r} (1 + z)^3\right)}{2 \left(\Omega_m (1 + z)^3 + \Omega_{r} (1 + z)^4 +\Omega_{\Lambda}\right)}.
\end{equation}
by using the relation $\dot{H} = -(1 +z) H(z) \frac{dH}{dz}$, and substituting the expression of $H(z)$ from Eq.~(\ref{eq:8}) into Eq.~(\ref{eq:25}). 

The deceleration parameter $ q > 0 $ indicates a decelerating expansion; $ q = 0$ corresponds to a constant expansion rate, and $ q < 0 $ indicates an accelerating expansion. In addition, the universe exhibits exponential expansion or de Sitter expansion when $ q =- 1 $, and superexponential expansion when $ q <- 1 $. It is also noted that the sign of $q$ already provides some key insight into how dynamically the Universe behaves.

Figure~\ref{fig:5a} illustrates the evolution of the deceleration parameter $ q(z) $ based on the observational dataset. The model predicts an early decelerating expansion phase ($ q > 0 $) during the matter-dominated era, followed by a transition to accelerated expansion ($ q < 0 $) as dark energy begins to dominate. This transition marks a significant shift in the expansion history of the universe. By examining the behavior of $ q(z) $, the study investigates how modifications to gravity can impact cosmic evolution. The redshift transition and the present value of the deceleration parameter are summarized in Table~\ref {Table 1}.

\subsection{Jerk Parameter}

To find the accelerated expansion of the universe, we study the higher-order time derivative of the scale factor. Jerk is a dimensionless parameter that includes the third derivative of the scale factor with respect to time. It is defined as
\begin{equation}\label{eq:27}
 j(t) = \frac{1}{H^3} \left(\frac{\dddot{a}}{a}\right).
 \end{equation}
The jerk parameter is extremely useful to distinguish a specific departure from the standard $\Lambda$CDM model. Also, it can be expressed in terms of the redshift $ z $ as
\begin{equation}\label{eq:28}
j(z) = (1 + z) \frac{dq(z)}{dz} + q(z) \left(1 + 2q(z)\right).
\end{equation}
Using Eqs.~(\ref{eq:26}) and~(\ref{eq:28}), we have the expression for the jerk parameter corresponding to our model is given by
\begin{equation}\label{eq:29}
    j(z) = \frac{\Omega_m (1 + z)^3 + 3 \Omega_{r} (1 + z)^4 + \Omega_{\Lambda}}
{\left(1 + z\right)^3 \left(\Omega_m + \Omega_{r} + \Omega_{r} z\right) + \Omega_{\Lambda}}.
\end{equation}
\begin{figure}
   \centering
    \subfloat[]{\label{fig:5a}\includegraphics[width=0.46\linewidth, height=0.35\linewidth]{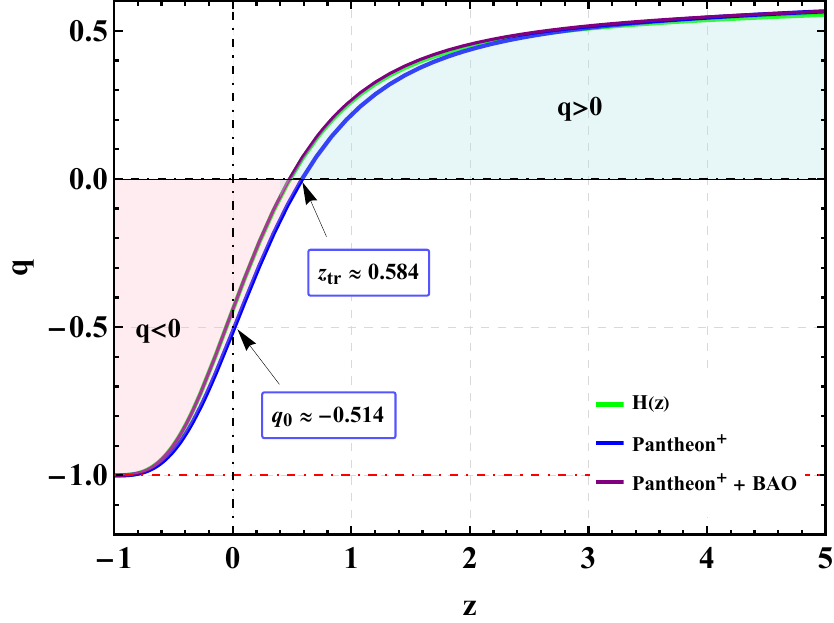}}\hfill
    \subfloat[]{\label{fig:5b}\includegraphics[width=0.46\linewidth, height=0.35\linewidth]{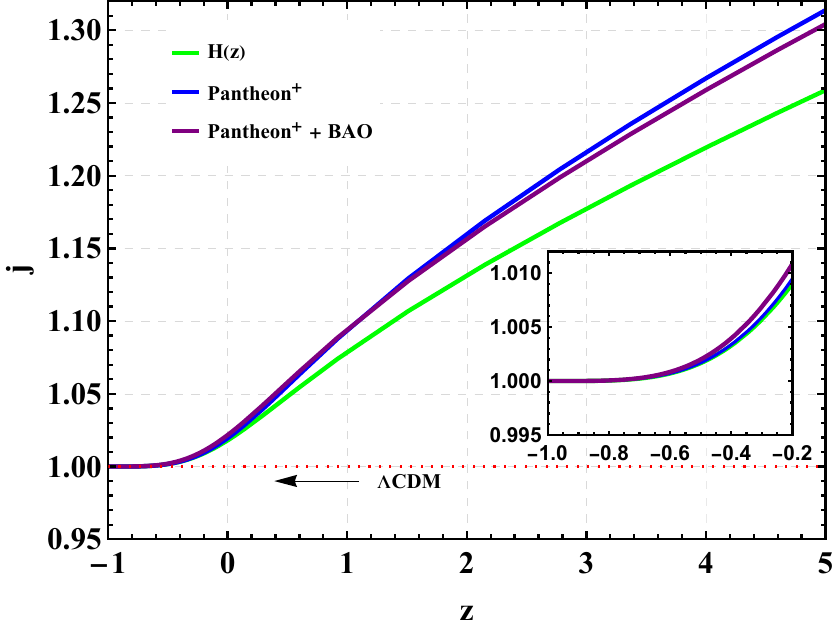}}\par
    \caption{Cosmic evolution of the deceleration parameter ($ q $) and the jerk parameter ($ j $) for all data compilations. }
    \label{Fig:5}
\end{figure}

A cosmic jerk is believed to be responsible for the universe switching from deceleration to acceleration in its expansion~\cite{Blandford:2004ah}. For instance, the $ \Lambda $CDM models is characterized by a constant jerk parameter, $ j(z) = 1 $~\cite{ Sahni:2002fz}.

As shown in Fig.~\ref{fig:5b}, the jerk parameter $ j(z) $ stays positive throughout the history of the universe. According to~\cite{Sahni:2002fz}, a constant value of $ j = 1 $ is a key feature of the standard $ \Lambda $CDM model. Interestingly, as seen from Eq.~(\ref{eq:29}), our model yields $ j(z) \approx 1 $ at the present epoch, indicating close agreement with the $ \Lambda$CDM model in the current epoch. However, its predictions may differ at earlier or later times.

\begin{figure}
    \centering
   \subfloat[]{\label{fig:6a}\includegraphics[width=0.46\linewidth, height=0.35\linewidth ]{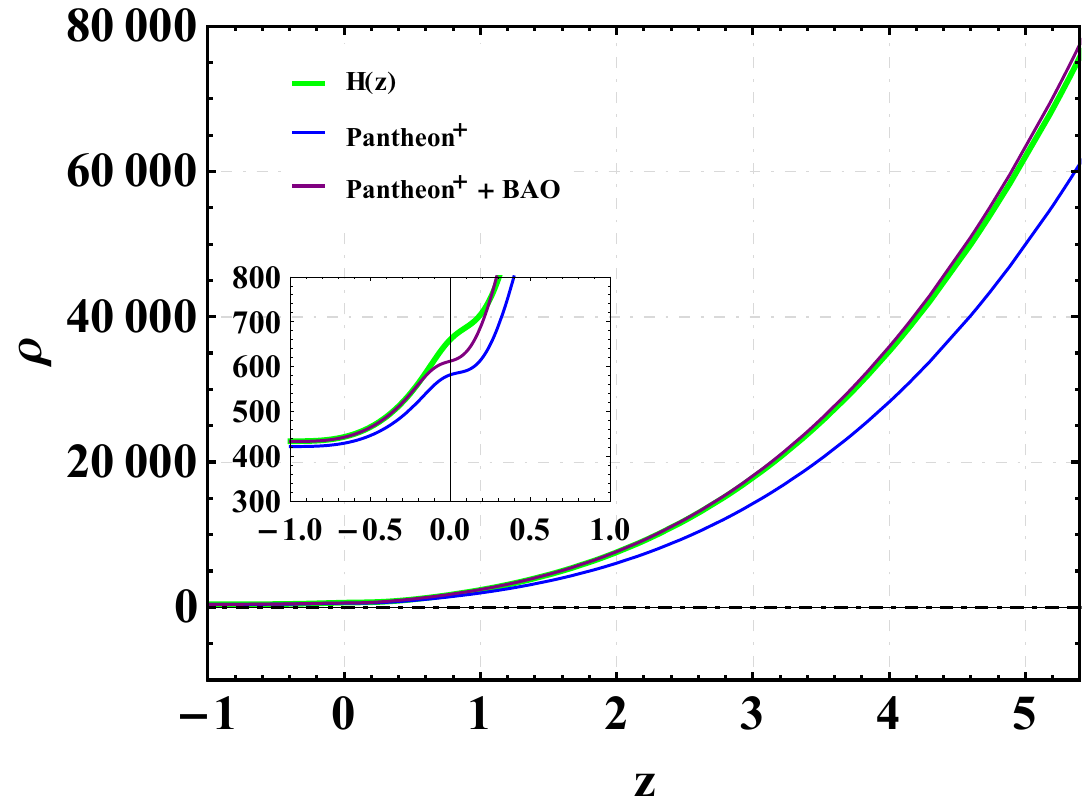}}\hfill
      \subfloat[]{\label{fig:6b}\includegraphics[width=0.46\linewidth, height=0.35\linewidth ]{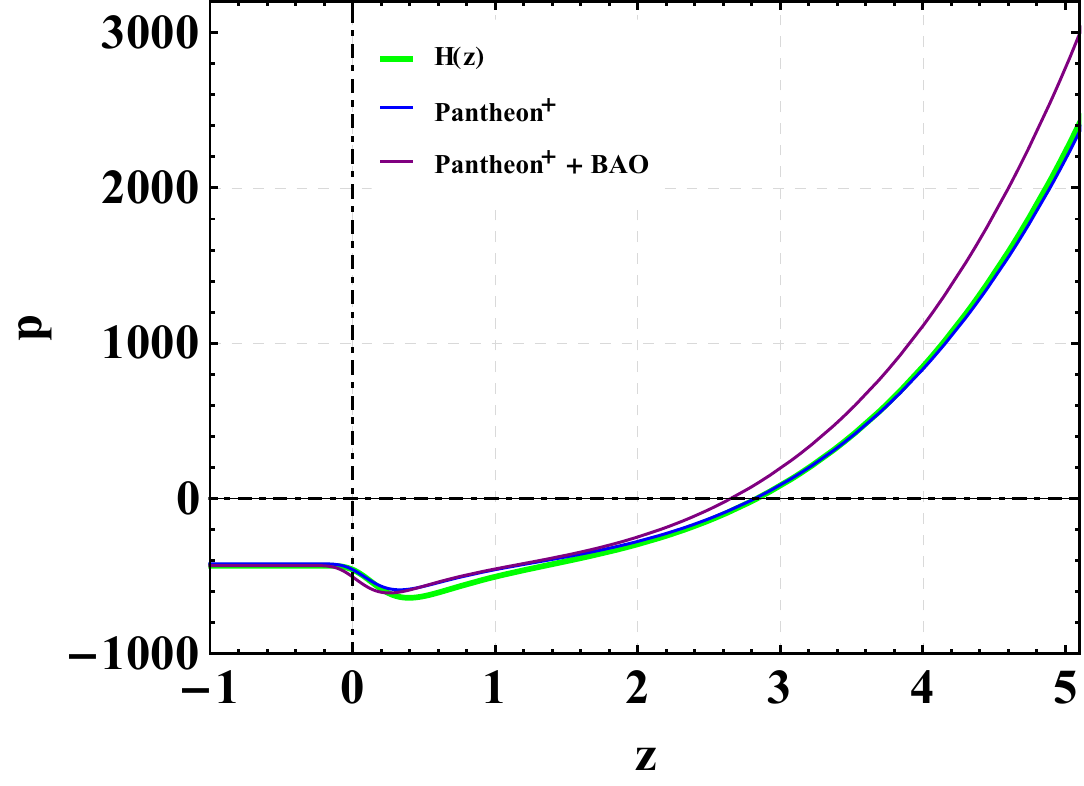}}\par
    \subfloat[]{\label{fig:6c}\includegraphics[width=0.46\linewidth, height=0.35\linewidth ]{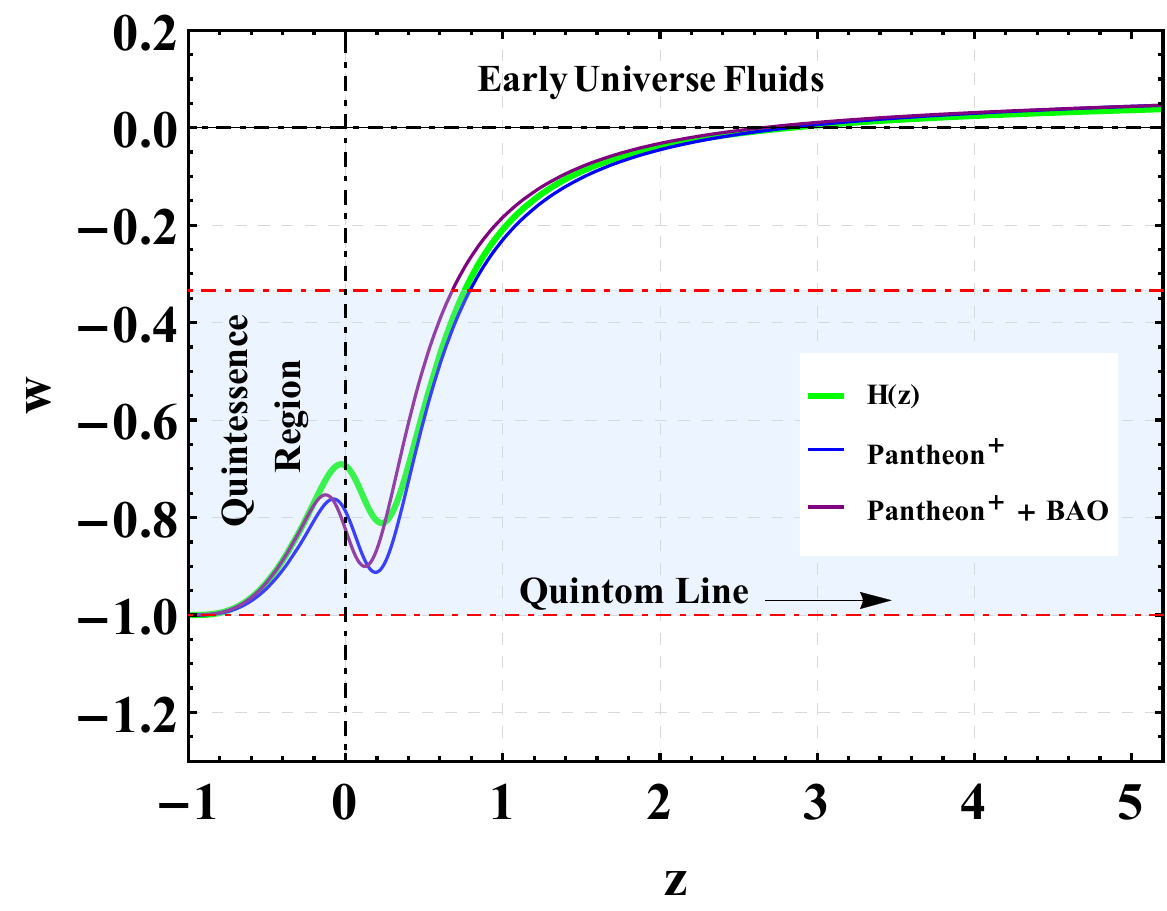}}
\caption{(a) The evolution of the energy density is positive ($ \rho >0 $), (b) The evolution of the matter pressure ($ p $) is negative, indicating an accelerating expansion of the model at late times, (c) The EoS parameter ($ w $) of the constrained model indicates the perfect fluid model on the redshift $ z>2.85 $, quintessence model in the redshift range $-1\leq z\leq0.82 $ for the observations Pantheon$^{+}$ and Pantheon$^{+}$ + BAO at late times.}
    \label{Fig:6}
\end{figure}

\subsection{The physical features of the Model}

We study how modifications to gravity affect key cosmological parameters, emphasizing physical interpretation over an exhaustive characterization of the full parameter space. Using best-fit values for the model parameters, we show that the considered models are consistent with current observations and thus represent plausible scenarios for cosmic evolution. These results contribute to improving our understanding of dark energy, dark matter, and the mechanisms driving the Universe’s accelerated expansion. By substituting the model expressions for $ \rho(z) $ and $ p(z) $ into Eqs.~(\ref{eq:5}) and (\ref{eq:6}), and fixing $ \zeta = 1 $, $ k = 0.16 $, and $ \alpha = 1.05 $, we obtain explicit redshift-dependent forms for the energy density and pressure.

\begin{align}
X(z) &= m_0(1+z)^3 + r_0(1+z)^4 + \lambda_0, \\
Q(z) &= \left(\frac{\alpha}{2H_0\sqrt{r_0}\,(1+z)^{3/2}}-\frac{\alpha\big(-2m_0+6r_0(1+z)-\lambda_0\big)} {4H_0 r_0^{3/2}\,(1+z)^4}\right)^{\!2}, \\
S(z) &= \operatorname{sech}^4\!\Bigg(\frac{\alpha\big(-2m_0+6r_0(1+z)-\lambda_0\big)}{12 H_0 r_0^{3/2}\,(1+z)^3}\Bigg),
\end{align}

\begin{align}
\rho(z) &= \frac{3H_0^2}{8\pi}\,X(z) - \frac{1}{2}\,H_0^2 k^2 (1+z)^2\,Q(z)\,X(z)\,S(z), \label{eq:rho} \\
p(z)    &= -\frac{H_0^2(1+z)^3}{8\pi}\big[\,3m_0 + 4r_0(1+z)\,\big]+ \frac{3H_0^2}{8\pi}\,X(z)
 - \frac{1}{2}\,H_0^2 k^2 (1+z)^2\,Q(z)\,X(z)\,S(z). \label{eq:p}
\end{align}

In Fig.~\ref{fig:6a}, the energy density $ \rho(z) $ remains positive and decreases monotonically as redshift $ z\to -1 $, consistent with the expected matter and radiation dominated Universe at the early epochs. The isotropic pressure $ p(z) $, in Fig.~\ref{fig:6b}, becomes positive at higher redshifts, reflecting the decelerated expansion in the past universe, and takes negative values at low redshift, indicating the late-time accelerated expansion.
 
In Fig.~\ref{fig:6c}, we observe the evolution of the EoS parameter $ w(z) $. his evolution suggests that the universe transitions from a radiation- and matter-dominated Universe at high redshift to a dark energy-dominated Universe at low redshift, consistent with a quintessence-type dark energy model. The EoS parameter $w $ approaches to $ -1 $ as $z\to-1$ and in the present epoch, this model exhibits the quintessence regime ($ -1 < w < -1/3 $).

\begin{figure}
    \centering
    \includegraphics[width=0.46\linewidth, height=0.35\linewidth]{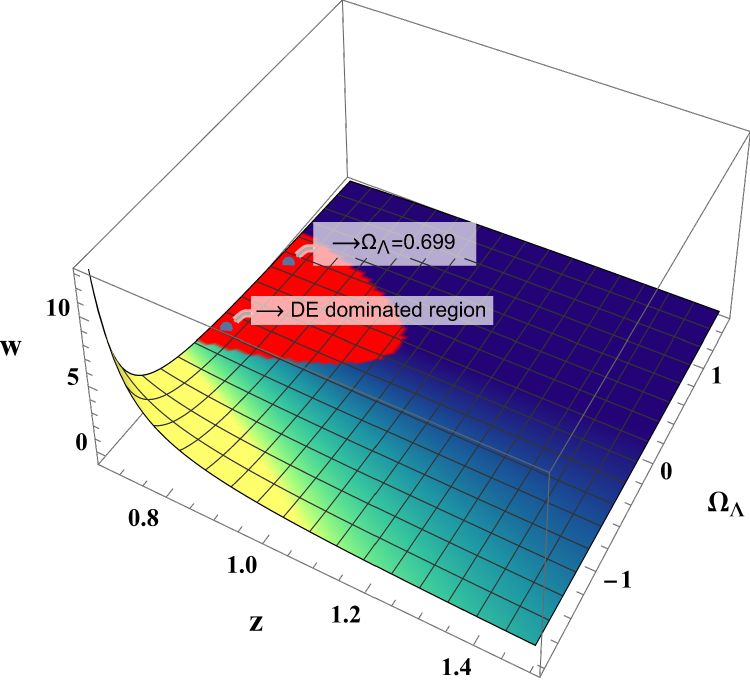}
   \caption{The cosmic evolution of the EoS in 3D of the constrained model for Pantheon $^{+}$ + BAO data and observe the consistency with the DESI DR2 results \cite{DESI:2025zgx} and the joint datasets results OHD+DES+BAO \cite{Shaily:2026xdk}.}
   \label{Fig:7d}
\end{figure}

In addition, the red zone in Fig.~\ref{Fig:7d} indicates that the model is dark energy-dominated, suggesting that quintessence dark energy is dominant at late times for the joint dataset Pantheon $^{+}$ + BAO.

\subsubsection{Statefinder Diagnostic}

The approach used to investigate the similarities between different classes of DE models and the $ \Lambda $CDM model is the statefinder diagnostic technique (SDT)~\cite{Singh:2015hva, Myrzakulov:2013owa, Rani:2014sia}. The statefinder diagnostic is a geometric parameter that is essential for understanding the evolution of the universe. It is proposed by Alam et al.~\cite{Alam:2003sc} and Sahni et al.~\cite{Sahni:2002fz}, defined as a pair of parameters ($ s, r $), and written in explicit forms as: 

\begin{equation}\label{eq:33}
\begin{split}
r &= \frac{\dddot{a}}{aH^3} = -q + 2q(1 + q) + (1 + z)\frac{dq}{dz}, \\
s &= \frac{r - 1}{3 \left(q - \frac{1}{2}\right)}, \quad \text{where } q \neq \frac{1}{2}.
\end{split}
\end{equation}
The statefinder diagnostic is a useful tool for distinguishing between different dark-energy models, as each cosmological scenario exhibits a unique evolutionary trajectory in the $ s–r $ plane. The $ \Lambda $CDM model corresponds to the point $ (s, r) = (0, 1) $, the Standard Cold Dark Matter (SCDM) model corresponds to $ (s, r) = (1, 1) $. The quintessence model corresponds to $ s > 0 $, the Chaplygin gas model yields $ s < 0 $ and $ r > 1 $, and the holographic dark-energy model is characterized by $ (s, r) = \left(\frac{2}{3},1 \right) $.

In addition to the $ (s, r) $ pair, a pair $ (q, r) $ is also commonly used, where the de Sitter state is defined by the point $ (q, r) = (-1, 1) $ and the $ \Lambda $CDM model is defined by the line $ r=1 $. The statefinder parameters for our model are given by

\begin{equation}\label{eq:34}
\begin{split}
r&= \frac{\Omega_m (1 + z)^3 + 3 \Omega_{r} (1 + z)^4 + \Omega_{\Lambda}}{\left(1 + z\right)^3 \left(\Omega_m + \Omega_{r} + \Omega_{r} z\right) + \Omega_{\Lambda}},\\
s&= \frac{4 \Omega_{r} (1 + z)^4}{3 \Omega_{r}(1 + z)^4 - 9 \Omega_{\Lambda}}.
\end{split}
\end{equation}

\begin{figure}
   \centering
    \subfloat[]{\label{fig:8a}\includegraphics[width=0.52\linewidth, height=0.37\linewidth]{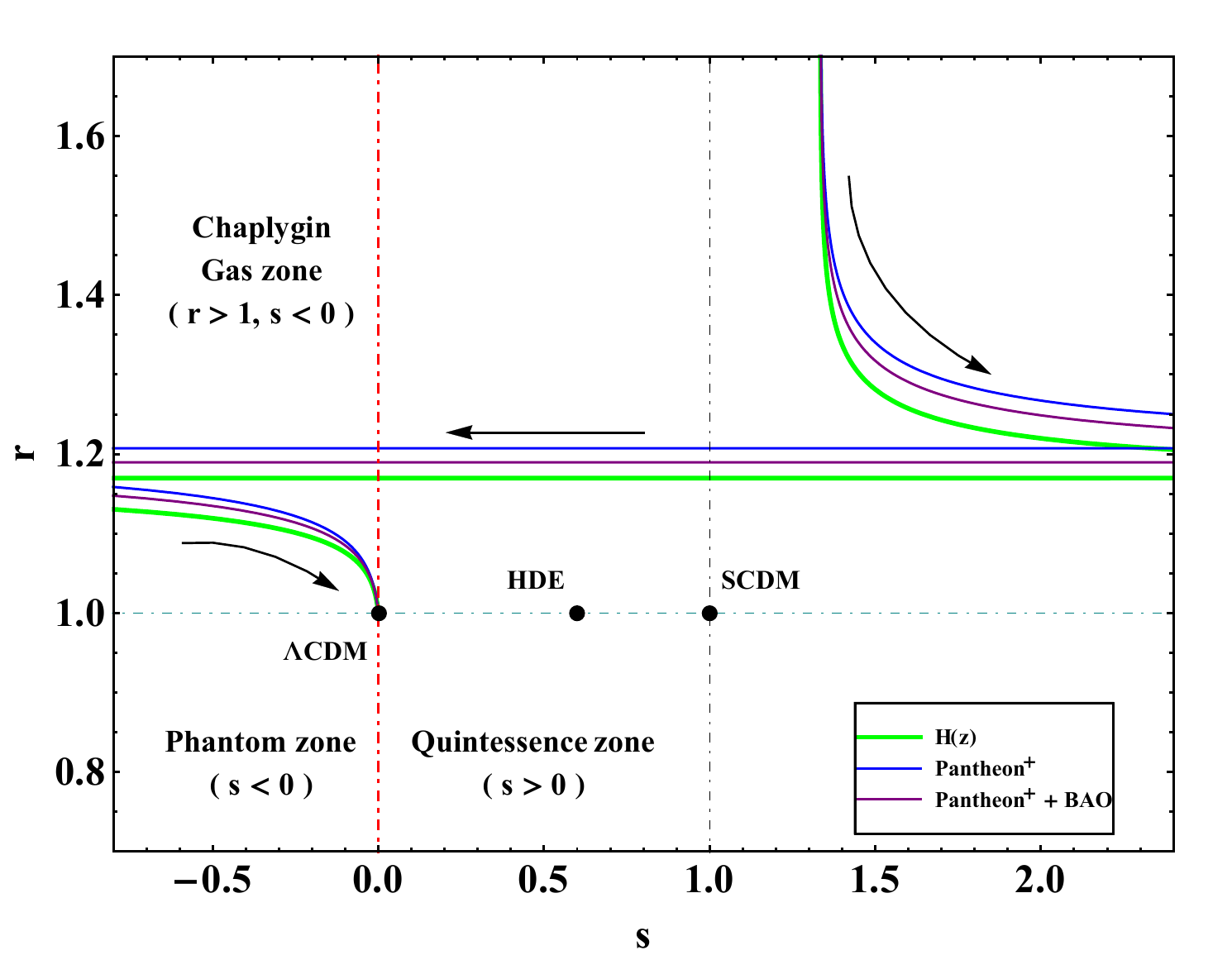}}\hfill
    \subfloat[]{\label{fig:8b}\includegraphics[width=0.45\linewidth, height=0.35\linewidth]{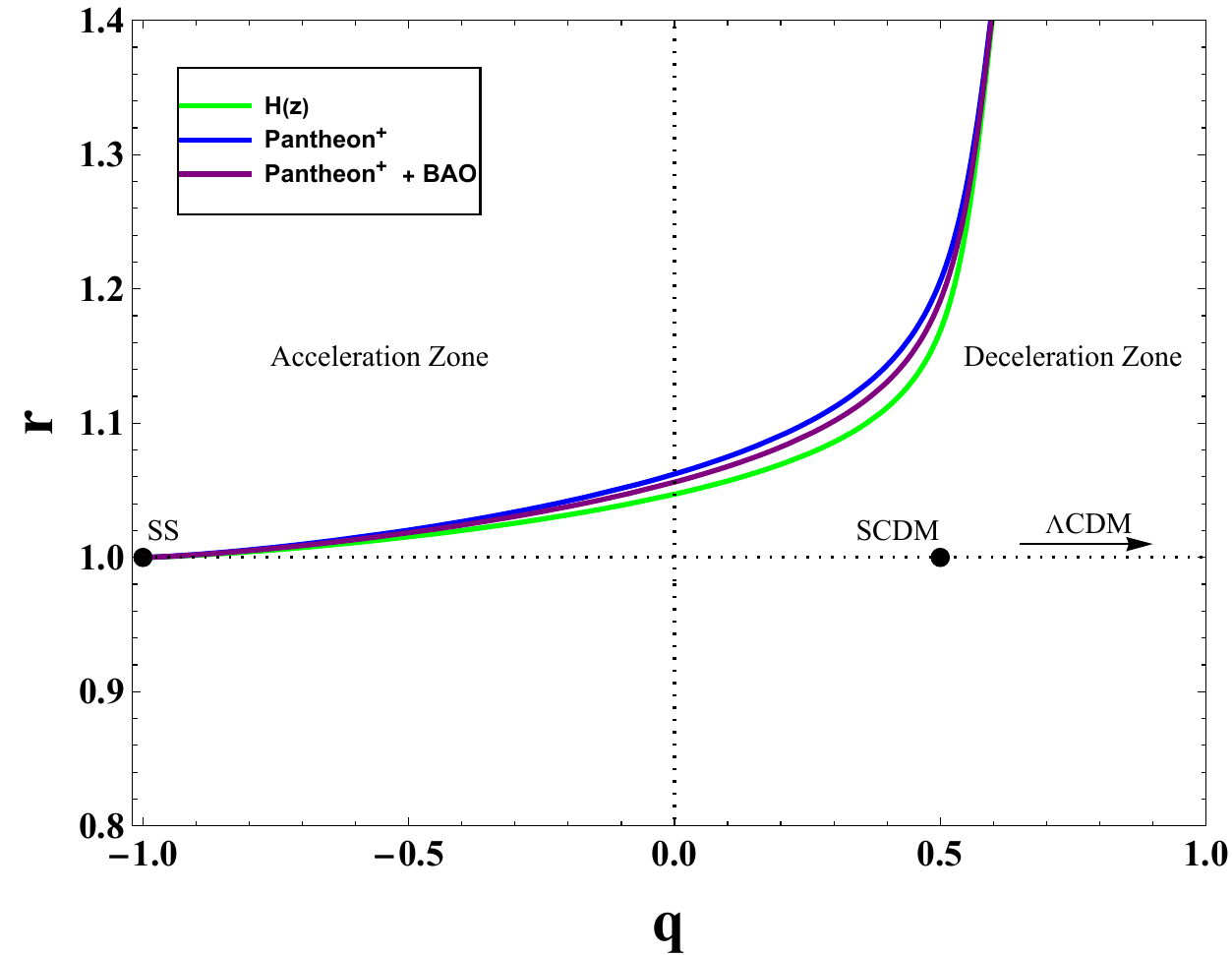}}\par
    \caption{ The cosmic evolution of the statefinder diagnostic pairs $\{s, r\}$ and $\{q, r\}$ of the constrained model. }
    \label{Fig:8}
\end{figure}

Figure~\ref{fig:8a} depicts the behavior of the statefinder pair $ (s,r) $. From Eq.~(\ref{eq:34}), we find $ r \to 1 $ and $ s \to 0 $ as redshift $ z \to -1 $. Now, using the best-fit value, the trajectories in the planes $s-r$ and $q-r$ are shown in Fig.~\ref{Fig:8}. The arrows in the plots indicate the evolution of trajectories over time, highlighting the Universe's phase transition and convergence towards the $ \Lambda $CDM model~\cite{Singh:2022eun, Nagpal:2018mpv}. 

The point $ (q, r) = \left( \frac{1}{2}, 1 \right) $ in the $q$–$r$ plane represents the Standard Cold Dark Matter (SCDM), while the point $ (q, r) = (-1, 1) $ corresponds to the Steady State (SS) model. The horizontal purple dotted line at $ r = 1 $ in Fig.~\ref{fig:8b} shows the path followed by the $ \Lambda $CDM model. As shown in Fig.~\ref{fig:8b}, the evolution of the cosmological models is illustrated. The model transitions from a decelerating state to an accelerating state and finally approaches a steady state at later times~\cite{Singh:2017qls}.

\begin{figure}
    \centering
    \subfloat[]{\label{fig:9a}\includegraphics[width=0.46\linewidth, height=0.35\linewidth]{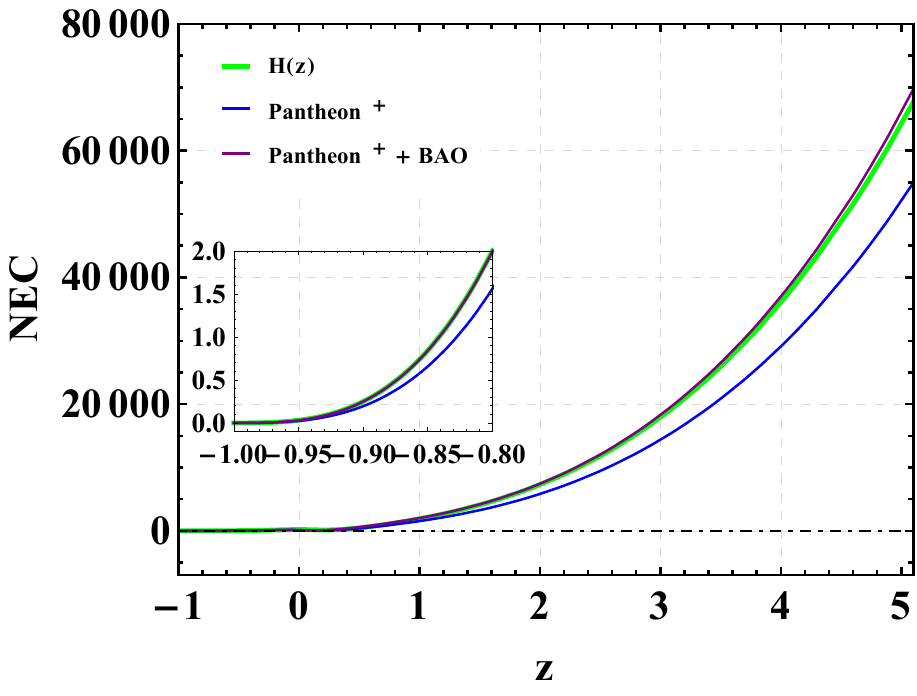}}\hfill
    \subfloat[]{\label{fig:9b}\includegraphics[width=0.46\linewidth, height=0.35\linewidth]{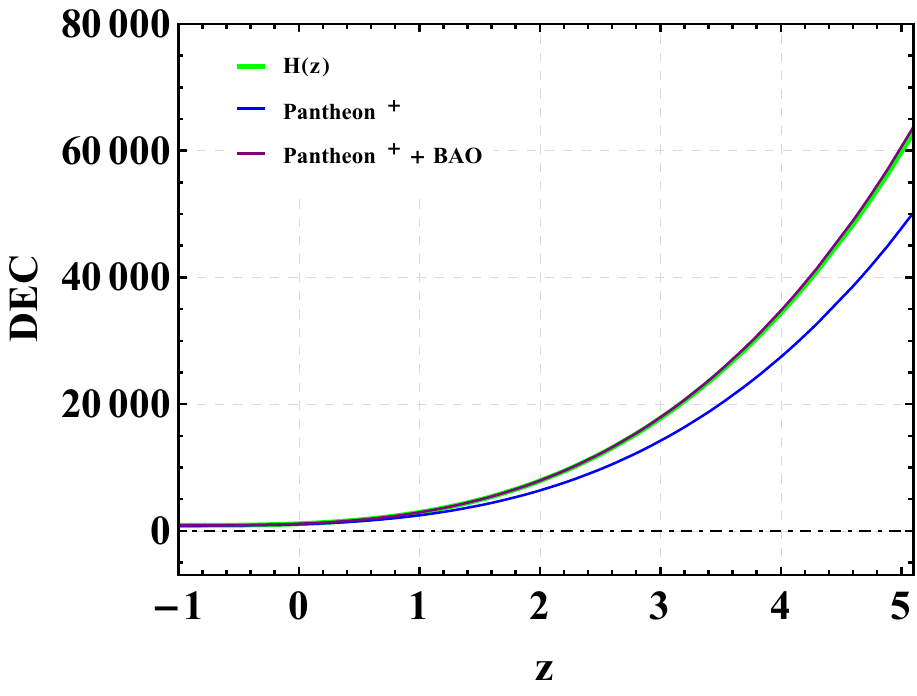}}\par
    \subfloat[]{\label{fig:9c}\includegraphics[width=0.46\linewidth, height=0.35\linewidth]{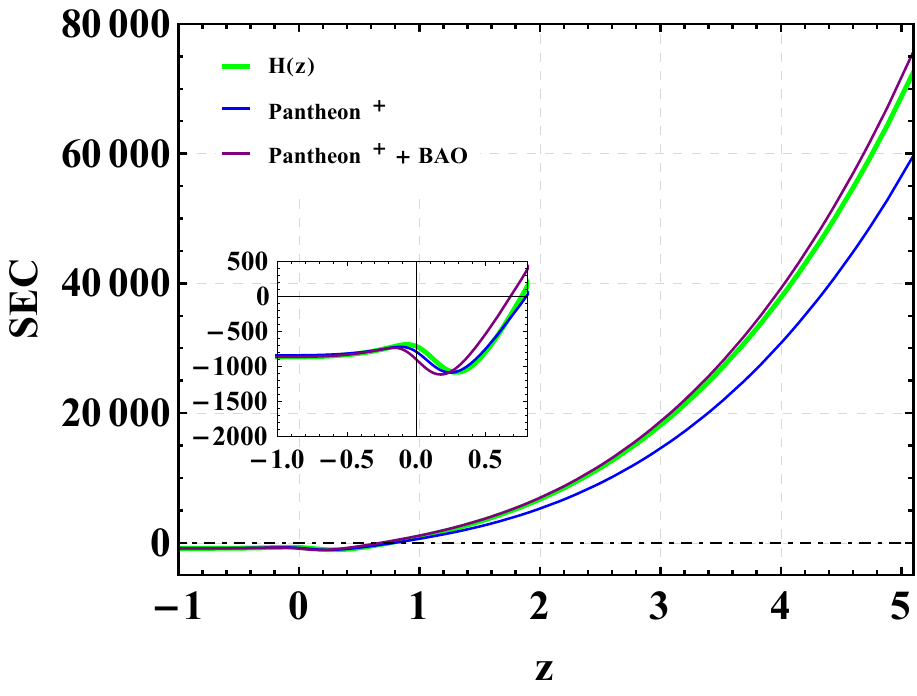}}
    \caption{(a) The NEC is positive during the evolution of the Universe, which signifies the certainty of singularities like the Big Bang in GR. (b) The DEC is positive during the evolution of the Universe, which states that the mass-energy density is always positive and the energy flows can never exceed the speed of light, ensuring the physical fidelity of the model, and is also a stricter condition than the WEC. (c) The violation of the SEC at late times states the rapid expansion of the Universe compelled by a matter creation field with negative pressure.}
    
    \label{Fig:9}
\end{figure}

\subsubsection{Energy Conditions}

In cosmology, the deceleration parameter (DP), energy density ($ \rho $), and the EoS parameter ($ w $), etc., are very crucial to study. However, another vital part of modern cosmology research focuses on the energy conditions originating from the Raychaudhuri equation~\cite{Raychaudhuri:1953yv, Singh:2022ptu, Raychaudhuri1957, Bhattacharyya:2021djv, Singh:2022ptu, Singh:2022eun, Das:2023umo, Panda:2023jwe}. These energy conditions serve to constrain the expansion of the Universe. We assume that the matter fields in the universe satisfy the energy conditions~\cite{Hawking:1973uf, Wald:1984rg}: null energy condition (NEC), weak energy condition (WEC), dominant energy condition (DEC), and strong energy condition (SEC) These conditions relate the energy density $ \rho $ and pressure $ p $ as follows: $ \rho + p \geq 0 $ and $ \rho \geq 0 $ (WEC), $ \rho + p \geq 0 $ (NEC), $ \rho \geq |p| $ and $ \rho \geq 0 $ (DEC), $ \rho + 3p \geq 0 $ (SEC).

As the Universe expands, the energy density decreases, which can be summarized as the violation of the NEC and explains the energy dilution phenomenon on cosmological scales. Similarly, the SEC violation indicates the accelerated expansion of the Universe~\cite{Visser:1999de}. Our interpretation of the energy conditions is presented in Fig.~\ref{Fig:9}. Here, we observe that from Fig.~\ref{fig:9a} and \ref{fig:9b}, we can see that the NEC and DEC conditions behave positively for the chosen model parameters. Since the WEC depends on both energy density and the NEC, it also holds throughout the entire redshift range. Interestingly, Fig.~\ref{fig:9c} shows that the SEC turns negative at lower redshifts, which represents the idea of cosmic acceleration~\cite{Capozziello:2013vna, Singh:2024urv, Singh:2022jue}. This pattern also appears in the deceleration parameter seen earlier in Fig.~\ref{fig:5a}.

The dominant energy condition (DEC) is a constraint on the stress-energy tensor in GR, ensuring that energy and momentum cannot flow faster than light. The evolution of the DEC is a physically inspired requirement that helps to define realistic matter distributions. Certainly, for any future illustration, the causal vector, which indicates the possible flow of energy, the energy flux measured in that direction cannot exceed the speed of light. When the DEC is violated, leading to scenarios where energy and momentum appear to travel faster than light, this has implications for causality. The physical significance of the DEC is crucial for ensuring that models of matter and gravity are physically plausible, and helps rule out exotic scenarios that might otherwise arise in general relativity calculations. In Fig.~\ref{Fig:7d}, we find that the energy and momentum do not travel faster than light in the evolution of the model, and satisfy all the physical laws in the model.

\begin{figure}
    \centering
    \includegraphics[width=0.46\linewidth, height=0.35\linewidth]{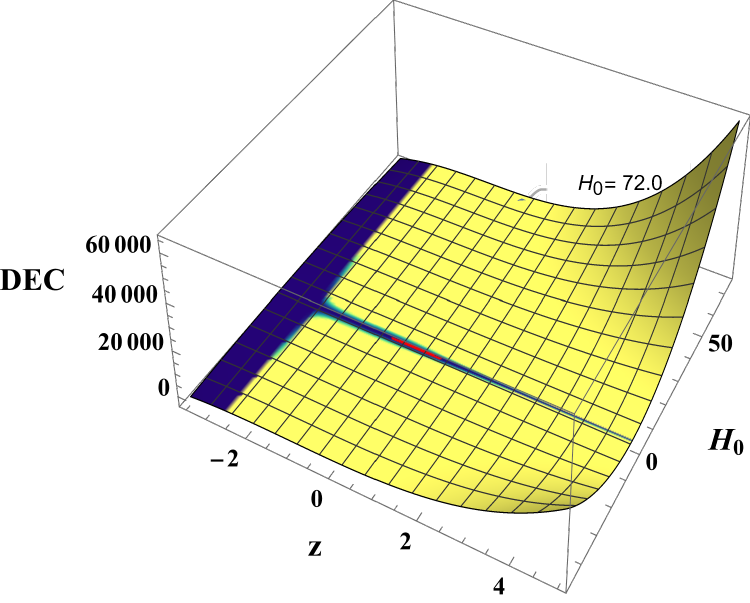}
    \caption{The cosmic analysis of DEC in 3D constrained model for Pantheon$^{+}$ + BAO data.}
    \label{Fig:10d}
\end{figure}

\section{Thawing and freezing analysis in the $ w-w'$ plane }\label{sec:6}

The $ w-w' $ diagnostic was first introduced by Caldwell and Linder~\cite{Caldwell:2005tm} as a tool to distinguish between different classes of dark energy models, which are broadly classified into thawing and freezing behaviors. In the framework of the $ w-w' $ plane, $ w' = \tfrac{dw}{d\ln a} $ measures the variation of $ w $ with respect to the logarithmic scale factor. Rather than presenting the lengthy explicit expression for $ w'(z) $ obtained by direct differentiation of $ w(z) $, we give here the compact, exact relation in terms of $ p(z) $ and $ \rho(z) $ given in Eqs.~\eqref{eq:rho} and \eqref{eq:p}, from which $ w'(z) $ can be straightforwardly derived.

\begin{figure}
   \centering
    \subfloat[]{\label{fig:11a}\includegraphics[width=0.47\linewidth, height=0.36\linewidth]{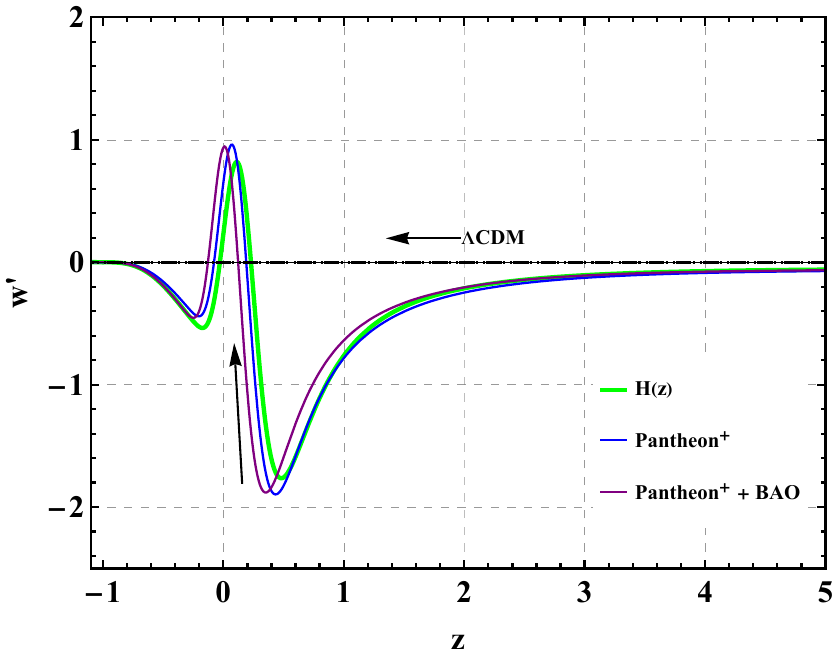}}\hfill
    \subfloat[]{\label{fig:11b}\includegraphics[width=0.47\linewidth, height=0.35\linewidth]{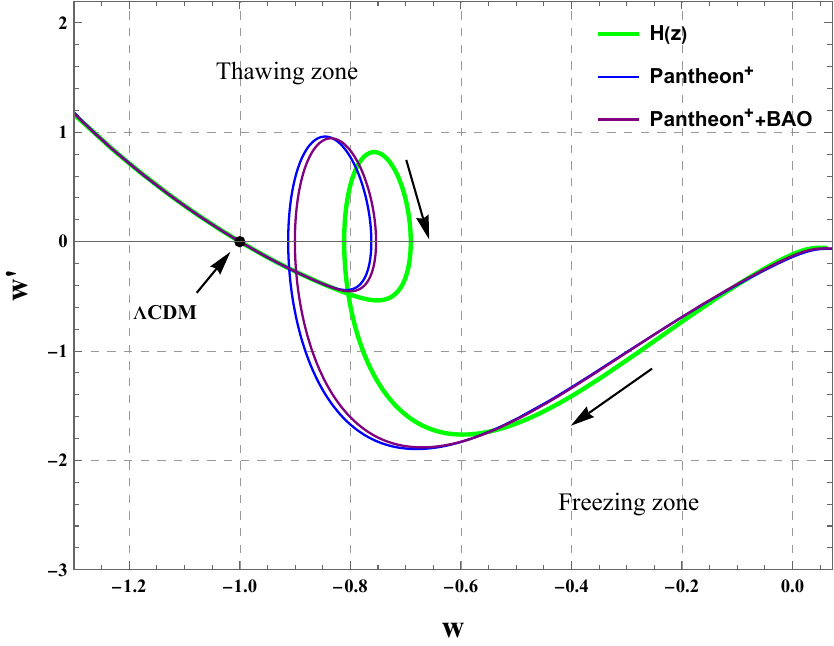}}\par
    \caption{The cosmic evolution of $ w'(z) $ and $ w-w' $ for all observational datasets.}
    \label{Fig:11}
\end{figure}

The trajectories in Fig.~\ref{fig:11a} reveal that $ w'(z) $ oscillates around the present epoch, with brief positive excursions, but eventually settles into negative values and asymptotically approaches zero at late times. This indicates an overall freezing behavior, where the rate of change of $ w $ slows as the Universe expands. Figure~\ref{fig:11b} shows the corresponding phase space trajectories in the $ w-w' $ plane. The point $ (-1,0) $ corresponds to the cosmological constant or $ \Lambda $CDM model, which acts as the late-time attractor. The trajectories in the figure show that the system does not approach this point monotonically but instead follows a spiral path with damped oscillations. When the curves pass through the region with $ w > -1 $ and $ w' > 0 $, the field exhibits thawing behavior, meaning that it temporarily moves away from the cosmological constant value. On the other hand, when the trajectories enter the region with $ w > -1 $ and $ w' < 0 $, the system shows freezing behavior, where the field evolves back toward $ \Lambda $CDM. Portions of the curves lying near $ w < -1 $ correspond to excursions into the phantom regime~\cite{Caldwell:1999ew, Carroll:2003st}, but these too are drawn back toward the attractor point. Overall, the figure demonstrates that the dark energy model alternates between thawing and freezing phases, producing oscillatory motion around $ (-1,0) $. Despite these intermediate deviations, the long-term trend is a stable convergence toward the cosmological constant, consistent with the observational bounds~\cite{Planck:2018vyg, DES:2021wwk}, highlighting that the model eventually mimics $ \Lambda $CDM at late times~\cite{Steinhardt:1999nw}.

\section{Concluding remarks}\label{sec:7}

\indent In this work, we perform a comprehensive analysis of the freezing quintessence framework in the late-time accelerating expansion. We consider the massless scalar field $ \mathcal{C} $-field known as the creation field that modifies Einstein’s field equations. Using the specific Hubble parameter as $ H(z) = H_0 \sqrt{(1 + z)^3 \Omega_m + (1 + z)^4 \Omega_{r} + \Omega_{\Lambda}} $, where $H_0$, $\Omega_m$, $\Omega_{r}$, and $\Omega_{\Lambda}$ , we obtain the cosmological solution. The model parameters are constrained using 77 $H(z)$ measurements, the Pantheon$^{+}$ supernovae dataset with 1701 data points, and their combination with BAO data. The summary of the best-fit values is presented in Table~\ref{Table 1}, which are consistent with observational data. Additionally, we examined the role of the $ \mathcal{C} $-field in the context of dark energy, assuming $ \mathcal{C}(t) = k \tanh(\alpha t) $. As shown in Fig.~\ref{fig:1}, the  $ \mathcal{C} $-field grows rapidly at early times and then saturates to a constant value, effectively behaving as a negative-pressure component that drives the late-time acceleration of the Universe. 

Moreover, we have observed the different physical performances of the model, such as\\ 
(i) the present values of the Hubble parameter are in good agreement with the Planck collaboration ($ H_0 = 67.4 \pm 0.5 ~km s^{-1} Mpc^{-1} $)~\cite{Planck:2018vyg}, \\
(ii)  According to the information-criteria analysis, both models fit the $ H(z) $ data well, whereas the Pantheon$^{+}$ and combined Pantheon$^{+}$+BAO datasets decisively favour the $ \Lambda $CDM model,\\
(iii) the deceleration parameter $ q $ transits from decelerating state to the accelerating state,\\
(iv) the jerk parameter converges to the $ \Lambda $CDM model in the late-times,\\
(v) the energy density $ \rho>0 $ during the cosmic evolution,\\
(vi) the matter pressure is negative in the late times,\\
(vii) The EoS parameter shows the perfect fluid behaviour on high redshift and converges to the quintessence model as $ z\to-1 $ according to the constrained values of the model parameters. At present, the values of the EoS for the datasets $ H(z) $, Pantheon$^{+}$, and Pantheon$^{+}$ + BAO are $ w_{0} = -0.69 $, $ w_{0} = -0.78 $, and $ w_{0} = -0.82 $, respectively.\\
(vii) The embedded dark energy region in the cosmic evolution of the EoS of the constrained model in 3D for Pantheon $^{+}$ + BAO data exhibits the quintessence behavior in later times (see Figs.~\ref{Fig:5}-\ref{Fig:7d}). 

The behavior of Statefinder Diagnostic pairs $\{s, r\} $ and $ \{q, r\} $ could be seen from Fig.~\ref{Fig:8}. The model starts with the deceleration era, passes through the quintessence and enters the Chaplygin regime, and finally seems to converge to the $ \Lambda $CDM model in the parametric space $ \{s, r\} $. Projection in the parametric space $ \{q, r\} $, the model passes from the deceleration region to the acceleration region, and approaches the de Sitter state. The energy conditions shown in Fig.~\ref{Fig:9} exhibit the positive nature of NEC and DEC conditions for all redshift values. However, there is one that the SEC has violated at the late-time epochs, corresponding to the observed accelerated expansion of the universe. In Fig.~\ref{Fig:10d}, we find that the energy and momentum do not travel faster than light in the evolution of the model, and satisfy all the physical laws in the model.

We further investigated dynamical stability through the $ w-w'$ phase trajectories, which confirm the alternating thawing behavior $ (w > -1,\,w' > 0) $ and freezing behavior $ (w > -1,\,w' < 0) $. Despite brief excursions, the system ultimately converges toward the $ \Lambda $CDM fixed point, confirming the model stability and consistency with late-time cosmic acceleration (see Fig.~\ref{Fig:11}). Finally, we conclude that the Hoyle-Narlikar gravity model is consistent with recent observational datasets as a quintessence-like accelerated expanding model of the universe.

\section*{Acknowledgement}
The authors express their gratitude to the Department of Mathematics, NSUT, New Delhi, India, and Bennett University, Greater Noida, India, for providing the necessary resources and support to facilitate the completion of this work. Joao R. L. Santos acknowledges support from CNPq (Grants 309494/2021-4 and 302190/2025-2), FAPESQ-PB (Grant 1356/2024), CAPES Finance Code 001, and Alexander von Humboldt-Stiftung Foundation.

\vspace{0.4cm}

\textbf{\noindent Data Availability Statement} No new data are created or analyzed in this study.

\vspace{0.4cm}

\textbf{\noindent Conflict of Interest} The authors declare that there are no conflicts of interest in the publication of this work.

\end{document}